\documentclass[preprints,article,accept,moreauthors,pdftex]{Definitions/mdpi}

\firstpage{1} 
\pubvolume{1}
\issuenum{1}
\articlenumber{0}
\pubyear{2024}
\copyrightyear{2024}
\datereceived{} 
\dateaccepted{} 
\datepublished{} 
\hreflink{https://doi.org/} 
\pdfoutput=1

\Title{EXACT MODEL OF GRAVITATIONAL WAVE AND PURE RADIATION}

\TitleCitation{Exact model of gravitational wave and pure radiation
}

\Author{
Konstantin E. Osetrin $^{1, 2, 3, 
\ddagger}$\orcidA, 
Vladimir Y. Epp $^{1,\ddagger}$ \orcidB, 
and 
Altair E. Filippov $^{1,\ddagger}$ \orcidC
}

\AuthorNames{Konstantin E. Osetrin, Vladimir Y. Epp and Altair E. Filippov}

\AuthorCitation{Osetrin, K.E.; Epp, V.Ya.; Filippov, A.E.}

\address{%
$^{1}$ \quad Tomsk State Pedagogical University; osetrin@tspu.edu.ru\\
$^{2}$ \quad Tomsk State University\\
$^{3}$ \quad Tomsk State University of Control Systems and Radioelectronics\\
}

\secondnote{These authors contributed equally to this work.}

\abstract{
An exact non-perturbative model of a gravitational wave with pure radiation is constructed. It is shown that the presence of dust matter in this model contradicts Einstein's field equations. Exact solution of Einstein equations for gravitational wave and pure radiation is obtained. The trajectories of propagation and characteristics of radiation are found. For the considered exact model of a gravitational wave, a retardet time equation for radiation is obtained. 
The obtained results are used to construct an exact model of gravitational wave and pure radiation for Bianchi type IV universe.
}

\keyword{
gravitational wave; pure radiation; light cone; retardet time of radiation; dust matter; Hamilton-Jacobi equation; Shapovalov spacetimes; Bianchi IV universe
} 

\MSC{83C10, 83C35}
\begin{document}
%

\section*{Introduction} \label{sec0}

The opening of the era of gravitational-wave astronomy (Nobel Prize in Physics, 2017) provided new tools for obtaining information about astrophysical objects and the universe as a whole \cite{PhysRevLett.116.061102,PhysRevX.9.031040,PhysRevX.11.021053}.
Studying the propagation of radiation and the motion of particles in a gravitational wave is an important task for gravitational-wave astronomy, since it can provide additional approaches and methods for recording and determining the characteristics of the gravitational-wave background \cite{ODINTSOV2024101562}.
Recently, this direction has received an additional experimental base in connection with the release of a number of publications on observational data on time delays of signals from pulsars when their radiation passes through the stochastic gravitational-wave cosmic background \cite{AandArefId0,Reardon_2023,Xu_2023}.

Usually, such models are considered in the perturbative approximation for weak gravitational perturbations against the background of the basic gravitational field
and are studied by numerical methods
\cite{Thorne:1980ru,Baumgarte:1998te,Blanchet:2013haa,MUKHANOV1992203,Ma19957}.
When constructing numerical models taking into account complex gravitational-wave signals, the availability of exact models of gravitational waves is of great importance, providing an exact mathematical basis and a transparent physical interpretation and allowing one to exactly calculate the influence of a gravitational wave on the propagation of radiation. Such exact models allow us to clarify the behavioral features of solutions and serve as a basis for debugging more complex numerical models of gravitational waves.

In this paper, we will consider the most general exact model of a gravitational wave, whose metric in a privileged coordinate system depends only on one wave variable, along which the space-time interval vanishes \cite{LandauEng1}. Along with the cosmological constant
in Einstein's field equations, we will consider such sources as pure radiation and dust matter. In this case, the pure radiation model can have a physical interpretation both as high-frequency electromagnetic radiation and as high-frequency gravitational radiation, as well as any other massless radiation against the background of a basic gravitational wave.

Exact models of gravitational waves are also of interest in the study of gravitational waves at the early stages of the universe's development, since observational data on the microwave cosmological background indicate its anisotropy \cite{Bennett2013}, and the generally accepted interpretation of this fact as a consequence of the kinematic effect during the motion of our galaxy also causes critical assessments \cite{Secrest_2021,refId0,10.1093/mnras/stad3706}. If the universe had anisotropy at the early stages of its development, then gravitational wave models based on Bianchi's anisotropic universe models and their influence on the formation of the microwave electromagnetic background of the universe may be of interest for research. The model of gravitational waves considered by us allows the existence of symmetries of Bianchi spaces \cite{OsetrinHomog2006} and can have joint exact solutions with the electromagnetic field \cite{Obukhov1983,Obukhov1986,Obukhov2021695,ObukhovSym14122595,Obukhov20202050186,ObukhovUniverse8040245,Obukhov2024_Einstein-Maxwell,ObukhovSym15030648} and, thus, can also model gravitational and electromagnetic waves in the early anisotropic universe. We have also previously constructed a number of exact models of similar gravitational waves \cite{Osetrin2022EPJP856,Osetrin2022894,Osetrin325205JPA_2023} for both Einstein's theory of gravity and modified theories of gravity \cite{Odintsov2007,Odintsov2011, Capozziello2011, Odintsov2017,Odintsov:2022hxu,Elizalde:2023rds,Nojiri:2023mbo,sym15091701,Odintsov:2024sbo}, including taking into account nonlinear terms in curvature in the field equations \cite{OsetrinSymmetry2021,Osetrin356Universe_2023}.

When studying the motion of test particles and the propagation of radiation in such exact models of gravitational waves, the Hamilton-Jacobi formalism is often used, which allows exactly integrating the equations of motion and obtaining the trajectories of particles and the trajectories of light propagation in a gravitational wave. The study of trajectories allows, among other things, to obtain the ratios of the delay of light signals during their propagation in gravitational waves \cite{OSETRIN2024169619}, which provides opportunities for experimental assessments of the characteristics of background gravitational waves.

From a general theoretical point of view, exact models of gravitational waves provide a general basis for deriving physical laws based on them, for example, for deriving Coulomb's law for the electric charge in a gravitational wave. From a technical point of view, non-perturbative exact models of gravitational waves are necessary to describe strong gravitational wave disturbances when perturbative methods do not work, for example, to describe gravitational waves at the early stages of the universe's development or near sources of strong gravitational wave disturbances. The delay of light signals in the stochastic gravitational wave background is currently being directly analyzed in the works of radio astronomers on long-term observations of pulsar signal delays.


\section{The gravitational wave exact model} 
\label{sec1}

Let us consider the most general model of a gravitational wave, the metric of which in a privileged coordinate system depends on only one wave variable. The space-time interval of the gravitational wave under consideration can be represented in the following general form
\cite{LandauEng1}:
\begin{equation}
{ds}^2=
2dx^0dx^1+g_{pq}(x^0)\Bigl(dx^p + f^p(x^0) dx^1\Bigr)\Bigl(dx^q + f^q(x^0) dx^1\Bigr)
\label{metric1}
.\end{equation}
Here $s$ is the space-time interval, indices \(p,q,r=2,3\). To write the metric (\ref{metric1}), a ''privileged'' coordinate system with the wave variable $x^0$ is used (along the wave variable, the space-time interval vanishes). The metric (\ref{metric1}) in the privileged coordinate system allows obtaining a complete integral for the Hamilton-Jacobi equation of test particles and for the eikonal equation
\cite{Shapovalov1978I,Shapovalov1978II,Shapovalov1979,Obukhov:2004yek}.

The equation for test particles in a gravitational field has the following form in the Hamilton-Jacobi formalism \cite{LandauEng1}:
\begin{equation}
g^{\alpha\beta}\frac{\partial S}{\partial x^\alpha}\frac{\partial S}{\partial x^\beta}
=
m^2 c^2
,\qquad
\alpha, \beta=0,1,2,3
\label{HJE}
,\end{equation}
where \(m\) is the mass, \(S\) is the action function of the test particle (unlike the interval, the action is denoted by a capital letter S), \(c\) is the speed of light, which we will set equal to unity in what follows.

As is known, the metric (\ref{metric1}) belongs to the class of Shapovalov wave metrics \cite{Osetrin2020Symmetry} and allows the existence of the so-called complete set of space-time symmetries (a complete set of Killing vector and tensor fields) that form a complete set of integrals of motion of the Hamilton-Jacobi equation for geodesics (\ref{HJE}), allowing one to obtain a complete integral of this equation in a privileged coordinate system \cite{Obukhov:2004yek}.

In a privileged coordinate system, the action function of a test particle can be represented in a separated form:
\begin{equation}
S (x^\alpha)=
{{\phi }_0}(x^0) + {{\lambda }_k}\,{x^k}
,\qquad i, j, k = 1,2,3.
\end{equation}
where \({\lambda }_k\) are constants (integrals of motion specified by initial or boundary conditions).

From the Hamilton-Jacobi equation (\ref{HJE}) for the metric (\ref{metric1}) we can write down the consequences that arise according to the method of separation of variables when the action function of a test particle in the considered privileged coordinate system has the form:
\begin{equation}
S=\phi_0(x^0)+\sum_k \lambda_k x^k
.\end{equation}
The Hamilton-Jacobi equation (\ref{HJE}) then gives the relation:
\begin{equation}
2\,\frac{d\phi_0}{dx^0}
\,
\Bigl(
\lambda_1-\sum_q \lambda_q f^q(x^0)
\Bigr)
=
m^2
-\sum_{p,q} \lambda_p \lambda_q\, g^{pq}(x^0)
,\end{equation}
where $\lambda_k$ are independent constants of motion of the test particle.

The action function of the test particle $S$ in the gravitational wave under consideration takes the following form:
\begin{equation}
S=
\sum_k \lambda_k x^k
+
\frac{1}{2}
\int
{
\frac{m^2
-\sum_{p,q}\lambda_p \lambda_q\, g^{pq}(x^0) }{\lambda_1-\sum_q \lambda_q f^q(x^0)}
\,dx^0
}
\label{FunctionS1}
.\end{equation}

The equations of the trajectory of the test particle in the Hamilton-Jacobi gravitational field formalism are written in the following form:
\begin{equation}
\frac{\partial S}{\partial \lambda_k}=\sigma_k
\label{ParticleTraject1}
,\end{equation}
where $\sigma_k$ are additional independent constant parameters of the particle trajectory, determined by the initial or boundary conditions of the test particle motion.

Substituting the obtained form of the function $S$ from (\ref{FunctionS1}) into the relations (\ref{ParticleTraject1}) we obtain the following general form of the equations of the trajectory of test particles (i.e.  the geodesic lines of space-time) in the considered gravitational wave in the privileged  coordinate system:
\begin{equation}
x^1
=
\sigma_1
+
\frac{1}{2}
\int{
\frac{m^2
-
\lambda_p \lambda_q\, g^{pq}(x^0)
}{
\left(
\lambda_1-
\lambda_q f^q(x^0)
\right)^2
}
\, dx^0
}
\label{ParticleTraject11}
,\end{equation}
\begin{equation}
x^p = \sigma_p - \frac{1}{2} \int { \frac{ f^p(x^0)\left( m^2 - \lambda_r \lambda_q\, g^{rq}(x^0) \right) }{ \left( \lambda_1 - \lambda_q f^q(x^0) \right)^2 } \, dx^0 } 
+
\int { \frac{ \lambda_q\, g^{pq}(x^0) }{ \lambda_1- \lambda_q f^q(x^0) } \, dx^0 } 
\label{ParticleTraject1p} 
.\end{equation} 
Here the wave variable $x^0$ plays the role of the parameter along the particle trajectories, determining the relationship between the variables $x^1$, $x^2$ and $x^3$ on the particle trajectories.

The proper time of a test particle $\tau$ can be defined via the particle action function by the relation (see \cite{LandauEng1}):
\begin{equation}
\tau
=
S/m
=
\sum_k {\tilde \lambda}_k x^k
+
\frac{1}{2}
\int
{
\frac{1
-
{\tilde \lambda}_p {\tilde \lambda}_q\, g^{pq}(x^0) }{{\tilde \lambda}_1
-
{\tilde \lambda}_q f^q(x^0)}
\,dx^0
}
,\qquad
{\tilde \lambda}_k= \lambda_k/m
\label{FunctionTau1}
.\end{equation}

Then we obtain from (\ref{FunctionTau1}) via the relations for $x^1$, $x^2$ and $x^2$ from the trajectory equations (\ref{ParticleTraject11})-(\ref{ParticleTraject1p}), the connection of the wave variable $x^0$ on the particle trajectory with the particle's proper time $\tau$.

Thus, via the equations (\ref{ParticleTraject11})-(\ref{FunctionTau1}), we obtain the trajectories of test particles in the usual notation $x^\alpha=x^\alpha(\tau)$, although the functions $x^\alpha(\tau)$ are generally specified not explicitly, but in parametric form.

The refinement of the form of the functions $g^{pq}(x^0)$ and $f^q(x^0)$, included in the metric of the wave model of space-time, arises from the field equations of the physical model and
the theory of gravity under consideration.

Note here that substituting the gravitational wave metric (\ref{metric1}) into Einstein's vacuum equations
\begin{equation}
R_{\alpha\beta}=0
,\end{equation}
where $R_{\alpha\beta}$ is the Ricci tensor,
leads to the following necessary conditions for the gravitational wave model under consideration (see \cite{LandauEng1}):
\begin{equation}
f^p(x^0)=0
.\end{equation}

\section{Pure radiation and dust matter in a gravitational wave} \label{sec2}

Let us consider the gravitational wave model for the case of Einstein's theory of gravity with field equations of the following form:
\begin{equation}
R_{\alpha\beta}-\frac{1}{2}R\, g_{\alpha\beta}
=
\Lambda\, g_{\alpha\beta}
+ \epsilon\, l_\alpha l_\beta
+\rho\, u_\alpha u_\beta
\label{EinstenEqs1}
,\end{equation}
where \(g_{\alpha\beta}\) is the spacetime metric for a gravitational wave of
signature \((+,-,-,-)\),
\(R_{\alpha\beta}\) is the Ricci tensor, \(R\) is the scalar curvature, \(\Lambda\) is the cosmological constant, \(\epsilon\) is the radiation energy density function (null dust), \(\rho\) is the dust matter mass density function, \(l_\alpha\) is the wave vector of pure radiation, \(u_\alpha\) is the four-velocity vector field of dust matter.

In this case, it is assumed that the following nomization conditions are satisfied:
\begin{equation}
\label{NullEqL}
g^{\alpha\beta} l_\alpha l_\beta=0
,\qquad
g^{\alpha\beta} u_\alpha u_\beta=1
.\end{equation}

The convolution of the field equations (\ref{EinstenEqs1}) then yields the following relation for the mass density of dust matter:
\begin{equation}
\rho= R-4\Lambda
\label{rho1}
.\end{equation}
Thus, if for the model under consideration and Einstein's field equations the scalar curvature \(R\) becomes constant, then the mass density of dust matter \(\rho\) will also be constant.

Calculating the scalar curvature \(R\) for the gravitational wave metric (\ref{metric1}) yields:
\begin{equation}
R =
-
\frac{
g
}{2}
\,
\left( {g^{33}}\,{({b^2})'}^2 + ({b^3})'\,\left(
{g^{22}}\,({b^3})'
-2\,{g^{23}}\,({b^2})'
\right) \right)
.\end{equation}

The components of the Ricci tensor \(R_{\alpha\beta}\) for the gravitational wave metric~(\ref{metric1}) are obtained in the following form:
\begin{equation} 
{R_{00}} = 
\frac{
1}{4\,g^2}
\,
\Bigl(
{{g}'}^2 + 2\,g^3\,\left( {({g^{23}})'}^2 - ({g^{22}})'\,({g^{33}})' \right)  - 2\,g\,{g}''
\Bigr)
\label{r00},\end{equation}  
\[
{R_{01}} = 
\frac{1}{4}
\,
\biggr[
-2 g \left( {g^{33}} {({b^2})'}^2 + {b^3}' \left( -2 {g^{23}} ({b^2})' + {g^{22}} ({b^3})' \right)  \right)  
\]
\[
\mbox{}
+ {b^3} 
\,
\biggl(
2 g \left( ({b^2})' ({g^{23}})'  - ({b^3})' ({g^{22}})'   \right)  
+ {g^{23}} \left( 3  {g}' ({b^2})' + 2 g ({b^2})'' \right)  
\]
\[
\mbox{}
- {g^{22}} \left( 3 {g}' ({b^3})' + 2 g ({b^3})'' \right) 
\biggr)
+ {b^2} 
\,
\biggl(
2 g \left( ({b^3})' ({g^{23}})' - ({b^2})' ({g^{33}})' \right) 
\]
\begin{equation} 
\mbox{}
- {g^{33}} \left( 3 {g}' ({b^2})' + 2 g ({b^2})'' \right)  + {g^{23}} \left( 3 {g}' ({b^3})' + 2 g ({b^3})'' \right)  
\biggr)
\biggr]
,\end{equation} 
\[
{R_{02}} = 
\frac{g}{2}
\Bigl(
{ ({b^2})' ({g^{33}})'}  - {({b^3})' ({g^{23}})'}  
+ { {g^{33}} ({b^2})''} - {{g^{23}} ({b^3})''}
 \Bigr)
\]
\begin{equation} 
\mbox{}
 +
 \frac{3{g}' }{4}
\Bigl(
{{g^{33}} ({b^2})'} - { {g^{23}}  ({b^3})'}
\Bigr)
,\end{equation}  
\[
{R_{03}} = 
\frac{g}{2}
\Bigl(
({b^3})' ({g^{22}})'-  ({b^2})' ({g^{23}})' -  {g^{23}} ({b^2})'' +  {g^{22}} ({b^3})''
 \Bigr)
 \]
\begin{equation} 
 \mbox{}
 +
 \frac{3{g}' }{4}
\Bigl(
{g^{22}} ({b^3})'
-
 {g^{23}} {g}' ({b^2})'
\Bigr)
,\end{equation}  
\begin{equation} 
{R_{11}} = 
\frac{- g^2 {
\Bigl(
{b^3} \left( {g^{23}} ({b^2})' - {g^{22}} ({b^3})' \right)  + {b^2} \left( 
{g^{23}} ({b^3})' 
- {g^{33}} ({b^2})'  
\right)  
\Bigr)
}^2 }{2}
,\end{equation}  
\[
{R_{12}} = 
\frac{g^2 
}{2}
\left( {g^{33}} ({b^2})' - {g^{23}} ({b^3})' \right)  
\times
\]
\begin{equation} 
\mbox{}
\times
\biggl(
{b^3} 
\left(
  {g^{22}} ({b^3})' 
 - {g^{23}} ({b^2})'  
 \right)  
 + {b^2} 
 \left( {g^{33}} ({b^2})' - {g^{23}} ({b^3})' \right)  
\biggr)
,\end{equation}  
\[
{R_{13}} = 
\frac{g^2 
}{2}
\left( {g^{23}} ({b^2})' - {g^{22}} ({b^3})' \right)  
\times
\]
\begin{equation} 
\mbox{}
\times
\biggl(
{b^3} \left( {g^{23}} ({b^2})' - {g^{22}} ({b^3})' \right)  
+ {b^2} \left(  {g^{23}} ({b^3})'- {g^{33}} ({b^2})'  \right)  
\biggr)
,\end{equation}  
\begin{equation} 
{R_{22}} = 
\frac{- g^2 {\left( {g^{33}} ({b^2})' - {g^{23}} ({b^3})' \right) }^2  }{2}
,\end{equation}  
\begin{equation} 
{R_{23}} = 
\frac{ g^2 \left( {g^{23}} ({b^2})' - {g^{22}} ({b^3})' \right)  \left(  {g^{33}} ({b^2})'   - {g^{23}} ({b^3})' \right)  }{2}
,\end{equation}  
\begin{equation} 
{R_{33}} = 
\frac{ -g^2 {\left( {g^{23}} ({b^2})' - {g^{22}} ({b^3})' \right) }^2  }{2}
.\end{equation}
Here \(g\) is the determinant of the gravitational wave metric:
\begin{equation}
g
=
\det{g_{\alpha\beta}}
= \frac{-1}{{g^{22}} {g^{33}}-\left({g^{23}}\right)^2 }
.\end{equation}

We choose some test particle as the base one, setting its mass \(m\) equal to unity and choosing its proper time \(\tau\) as a parameter along the trajectory. Then the components of the 4-velocity of particles in the used privileged coordinate system according to the equations of trajectories
(\ref{ParticleTraject11})-(\ref{FunctionTau1}) can be written as follows:
\begin{equation}
u_\alpha=g_{\alpha\beta}\,u^\beta=g_{\alpha\beta}\,\frac{dx^\beta}{d\tau}=\Bigl\{u_0(x^0),\lambda_1,\lambda_2,\lambda_3 \Bigr\}
.\end{equation}
\begin{equation}
u_0=
\frac{1
-\sum_{p,q}\lambda_p \lambda_q\, g^{pq}(x^0) }{2\left(\lambda_1-\sum_q \lambda_q f^q(x^0)\right)}
,\qquad
p,q,r=2,3
.\end{equation}

Using the relation (\ref{rho1}) from the field equations
with the components of the Ricci tensor \(R_{22}\), \(R_{23}\) and \(R_{33}\)
we obtain the following form of these equations:
\begin{equation}
{{{({b^3})'}^2}= {\frac{-2\,\left( \varepsilon \,{{l_2}}^2 + \rho \,{{{\lambda }_2}}^2 \right) }{g} - \left( 10\,\Lambda + 3\,\rho \right) \,{g^{33}}}} 
\label{eq71} 
,\end{equation} 
\begin{equation} {{({b^2})'\,({b^3})'}={\frac{2\,\left( \varepsilon \,{l_2}\,{l_3} + \rho \,{{\lambda }_2}\,{{\lambda }_3} \right) }{g} - \left( 10\,\Lambda + 3\,\rho \right) \,{g^{23}}}} 
\label{eq72} 
,\end{equation} 
\begin{equation} {{{({b^2})'}^2}= {\frac{-2\,\left( \varepsilon \,{{l_3}}^ 2 + \rho \,{{{\lambda }_3}}^2 \right) }{g} - \left( 10\,\Lambda + 3\,\rho \right) \,{g^{22}}}} 
\label{eq73} 
.\end{equation} 

We obtain the compatibility condition for the equations (\ref{eq71})-(\ref{eq73}) in the following form: \begin{equation} \left( 2\,\Lambda + \rho \right) \,\left( 10\,\Lambda + 3\,\rho \right) \,g + 4\,\varepsilon \,\rho \,{\left( {l_3}{{\lambda }_2} - {l_2}{{\lambda }_3} \right) } ^2 = 0
.\end{equation}

From the field equations with components $R_{11}$, $R_{12}$, and $R_{13}$, we obtain the following relations:
\begin{equation} \varepsilon\,l_1\,(l_1+l_p\,b^p)+\rho\,\lambda_1\,(\lambda_1+\lambda_p\,b^p)=0 \label{eq75}
,\end{equation}
\begin{equation} \varepsilon\,l_2\,(l_1+l_p\,b^p)+\rho\,\lambda_2\,(\lambda_1+\lambda_p\,b^p)=0 \label{eq76}
,\end{equation}
\begin{equation} \varepsilon\,l_3\,(l_1+l_p\,b^p)+\rho\,\lambda_3\,(\lambda_1+\lambda_p\,b^p)=0 \label{eq77}
.\end{equation}
Assuming that $\rho\neq 0$, from equations (\ref{eq75})-(\ref{eq77}) we obtain corollaries of the form:
\begin{equation}
{l_i}\,{{\lambda }_j} - {l_j}\,{{\lambda }_i} =0,
\qquad
\left( 2\,\Lambda + \rho \right) \,\left( 10\,\Lambda + 3\,\rho \right)=0
\label{eq37}
.\end{equation}
Equations (\ref{eq75})-(\ref{eq77}), taking into account the relation $l_k=\kappa \lambda_k$ following from (\ref{eq37}) (here \(\kappa\) is an arbitrary constant) lead to the following condition:
\begin{equation}
\lambda_k\,(\lambda_1+\lambda_p\,b^p)(\varepsilon \kappa^2+\rho)=0
.\end{equation}
Since all \(\lambda_k\) cannot vanish simultaneously, we obtain a corollary of the form:
\begin{equation}
\varepsilon \kappa^2+\rho=0
\label{rho2}
.\end{equation}
Putting the condition (\ref{rho2}) into the equations (\ref{eq71}-\ref{eq73}) we obtain: \begin{equation} {({b^3})'}^2= - \left( 10\,\Lambda + 3\,\rho \right) \,{g^{33}} \label{eq81},\end{equation} \begin{equation} {( {b^2})'\,({b^3})'}= - \left( 10\,\Lambda + 3\,\rho \right) \,{g^{23}} \label{eq82},\end{equation} \begin{equation} {{({b^2})'}^2}= - \left( 10\,\Lambda + 3\,\rho \right) \,{g^{22}}
\label{eq83}
,\end{equation}
then as a result of $g^{22}g^{33}-{g^{23}}^2=\det g^{\alpha\beta}\neq 0$,
we obtain equations of the form:
\begin{equation}
10\,\Lambda + 3\,\rho =0\quad \rightarrow\quad (b^p)'=0
.\end{equation}

The obtained conditions of compatibility of the field equations lead to a contradiction with the equation (\ref{rho1}). Thus, we have arrived at a contradiction, from which it follows that in the gravitational wave model under consideration, dust matter in the given form cannot exist and \(\rho=0\).

\section{Exact solution of Einstein's equations with pure radiation} \label{sec3}

Let us now consider the case when in the considered model of gravitational wave dust matter is absent and \(\rho=0\), but pure radiation with wave vector \(l^\alpha\) and energy density \(\varepsilon\) is preserved. Then the reduction of the field equations taking into account the presence of pure radiation yields:
\begin{equation}
\rho=\Lambda=b^p=0
\label{Solution1}
,\end{equation}
\begin{equation}
{ds}^2=
2\,dx^0dx^1+g_{pq}(x^0)\,dx^p dx^q
,\qquad
g=\det{g_{\alpha\beta}}=-\det{g_{pq}}
,\quad
p,q=2,3
\label{metric2}
,\end{equation}
\begin{equation}
l_\alpha=(l_0,0,0,0),
\qquad
l^\alpha=(0,l_0,0,0)
\label{Solution2}
,\end{equation}
\begin{equation}
\varepsilon\,l_0{}^2=\frac{{g'}^2-2gg''+2g^3\left( {(g^{23})'}^2-{(g^{22})}'{(g^{33})}'\right)}{4g^2}
\label {Solution3}
.\end{equation}
In this case, the scalar curvature of the space-time of the gravitational wave \(R\) and the cosmological constant \(\Lambda\) vanish.

This exact solution for the gravitational wave and pure radiation (\ref{Solution1})-(\ref{Solution3}) can have several variants of physical interpretation depending on the formulation of the problem.

First, pure radiation can be considered as an external source of energy
(for example, external electromagnetic radiation or a flow of other massless particles).
In this case, the external pure radiation is assumed to be specified through the wave vector \(l_\alpha\) and the radiation energy density \(\varepsilon\).
Then its presence gives an additional constraint on the three functions \(g^{22}\), \(g^{33}\) and \(g^{23}\) - the remaining components of the gravitational wave metric in the form of equation (\ref{Solution3}). In this case, the obtained solution exists under the condition that the external radiation has a wave vector of the form (\ref{Solution2}). Thus, the characteristics of the external pure radiation (\(l_\alpha\) and \(\varepsilon\)) determine both the choice of the used privileged coordinate system and the gravitational wave itself.

Secondly, pure radiation can be considered as high-frequency gravitational radiation against the background of a slowly changing basic metric of the gravitational wave, determined by the components of the metric remaining after the reduction \(g^{22}\), \(g^{33}\) and \(g^{23}\), which in this formulation of the problem are considered arbitrary functions. Equation (\ref{Solution3}) then determines the intensity of the high-frequency part of the gravitational wave, and the wave vector (\ref{Solution2}) determines the direction of this high-frequency gravitational radiation against the background of the slowly changing part of the gravitational wave with the metric  functions \(g^{pq}(x^0)\).

Thirdly, pure radiation can be considered as a flow of massless particles generated by a gravitational wave
due to the energy of the wave (for example, the generation of an electromagnetic wave), i.e. the gravitational wave in this case will ''glow''. The direction of the radiation generated by the gravitational wave will be determined by the wave vector (\ref {Solution2}), and the intensity of this radiation 
will be determined by equation (\ref{Solution3}) according to the metric of the gravitational wave.

\section{Light cone and radiation delay in a gravitational wave} \label{sec4}

The propagation of light signals in a gravitational wave is determined by the eikonal equation
\begin{equation}
g^{\alpha\beta}\frac{\partial \Psi}{\partial x^\alpha}\frac{\partial \Psi}{\partial x^\beta}
=
0
\label{eikonalEq}
,\end{equation}
where \(\Psi\) is the eikonal function that determines the propagation front of the light signal.

By analogy with the previous calculations for the trajectories of test particles, the solution of the eikonal equation (\ref{eikonalEq}) by the method of separation of variables leads to the following form of equations for the trajectories of light rays in a gravitational wave with metric (\ref{metric1}):
\begin{equation}
x^1
=
\gamma_1
-
\frac{1}{2}
\int{ 
\frac{
\sum_{p,q}{k}_p {k}_q\, g^{pq}(x^0)
}{
\left(
{k}_1-\sum_q {k}_q f^q(x^0)
\right)^2
}
}\,dx^0
,\qquad
p,q, r,s=2,3.
\label{LightTraject11} 
\end{equation} 
\begin{equation} 
x^p = \gamma_p 
+ \frac{1}{2} 
\int{
\frac{ f^p(x^0) \sum_{r,q}{k}_r {k}_q\, g^{rq}(x^0) }{ \left( {k}_1-\sum_q { k}_q f^q(x^0) \right)^2 } 
}\,dx^0
+ 
\int{ 
\frac{ \sum_{q} {k}_q\, g^{pq}(x^0) }{ {k}_1-\sum_q {k}_q f^q(x^0)}
}\,dx^0
\label{LightTraject1p}
,\end{equation}
where independent parameters ${k}_i$ and $\gamma_i$ are determined by the initial or boundary conditions of radiation propagation in a gravitational wave, the wave variable $x^0$ plays the role of a parameter along the trajectory of light rays propagation.

By selecting and fixing one of the world points \(x^\alpha_{(D)}\), through which all possible trajectories of light rays (\ref{LightTraject11})-(\ref{LightTraject1p}) pass, we obtain the equation of the light ''cone'' for this world point, along which light signals in a gravitational wave propagate, which can be detected by an observer at this point.

If we now additionally fix another world point \(x^\alpha_{(S)}\), in which the radiation source is located, the light signal of which is detected by an observer at the point \(x^\alpha_{(D)}\), then we obtain light trajectories that connect the world points of the source and the signal detector. These relations define boundary conditions that allow us to express the parameters of the light trajectory \(\gamma_k\) and \({k}_q/k_1\) through the coordinates of the world points of the source and detector of the signal \(x^\alpha_{(S)}\) and \(x^\alpha_{(D)}\), and also to find the equation of the delay of the light signal, which relates the coordinates of the world points of the source and detector. 
For calculation details, see the appendix (\ref{AppendixA}).
Such a "delay" relation in reference systems with an explicitly distinguished time variable gives the connection between the moment of emission and the moment of detection of the light signal during its propagation in a gravitational wave (see \cite{OSETRIN2024169619}).

Solving the boundary condition equations for the parameters of the light signal trajectory in the exact gravitational wave model with the metric (\ref{metric2}) and with the conditions (\ref{Solution1})-(\ref{Solution3}), we obtain expressions for the parameters \(\gamma_p\) of the form:
\begin{equation}
\gamma_p=
\sum_q
\Bigl[ 
\left(
I-{G_{\scriptscriptstyle D}}G_{\scriptscriptstyle S}^{-1}
\right)^{-1}
\Bigr]_{pq}
\left(
x^q_{\scriptscriptstyle D}
-
\sum_{r}
\Bigl[
{G_{\scriptscriptstyle D}}G_{\scriptscriptstyle S}^{-1}
\Bigr]_{qr}
x^r_{\scriptscriptstyle S}
\right)
\label{gammaPend1}
,\end{equation}
\begin{equation}
G^{pq}(x^0)=
\int{g^{pq}(x^0)}\,dx^0
,\qquad
G_{\scriptscriptstyle D}=G^{pq}(x^0_{\scriptscriptstyle D})
,\qquad
G_{\scriptscriptstyle S}=G^{pq}(x^0_{\scriptscriptstyle S})
\label{functionG}
.\end{equation}
where $G(x^0)$ is the matrix from the relations (\ref{functionG}), $I$ is the identity matrix
(inverse matrices and matrix products are used in expressions).
The subscript D means that the quantity refers to the detector, and the subscript S means that the quantity refers to the signal source.

The parameters of the radiation trajectory $k_p/k_1$ 
we obtain from the boundary conditions in the following form:
\[
\frac{k_p}{k_1}
=
\sum_{q}
\Bigl[\left({G_{\scriptscriptstyle D}}-G_{\scriptscriptstyle S}^{-1}\right)^{-1}\Bigr]_{pq}
x^q_{\scriptscriptstyle D}
\]
\begin{equation} 
\mbox{}
+
\sum_{q}
\left(
\left[G_{\scriptscriptstyle S}^{-1}\right]_{pq}
-
\Bigl[\left({G_{\scriptscriptstyle S}}-{G_{\scriptscriptstyle S}}G_{\scriptscriptstyle D}^{-1}G_{\scriptscriptstyle S}^{-1}\right)^{-1}\Bigr]_{pq}
\right)
x^q_{\scriptscriptstyle S}
\label{kappaPend1}
.\end{equation} 

For the light ray trajectory parameter $\gamma_1$ we obtain the following expression:
$$
\gamma_1
=
x^1_{\scriptscriptstyle S}
+
\frac{1}{2}
\sum_{p,q}
x^p_{\scriptscriptstyle S}
\left[G_{\scriptscriptstyle S}^{-1}\right]_{pq} x^q_{\scriptscriptstyle S}
$$
$$
\mbox{}
-
\frac{1}{2}
\sum_{p,q}
\left(
x^p_{\scriptscriptstyle S}
-
\sum_{r}
\Bigl[
{G_{\scriptscriptstyle D}}G_{\scriptscriptstyle S}^{-1}
\Bigr]_{pr}
x^r_{\scriptscriptstyle S}
\right)
\sum_r
\Bigl[
\left(
{G_{\scriptscriptstyle S}}-{G_{\scriptscriptstyle D}}
\right)^{-1}\Bigr]_{rp}x^r_{\scriptscriptstyle S}
$$
$$
\mbox{}
-
\frac{1}{2}
\sum_{p,q}
x^p_{\scriptscriptstyle S}
\left[\left({G_{\scriptscriptstyle S}}-{G_{\scriptscriptstyle D}}\right)^{-1}\right]_{pq} 
\left(
x^q_{\scriptscriptstyle D}
-
\sum_{r}
\Bigl[
{G_{\scriptscriptstyle D}}G_{\scriptscriptstyle S}^{-1}
\Bigr]_{qr}
x^r_{\scriptscriptstyle S}
\right)
$$
$$
\mbox{}
-
\frac{1}{2}
\sum_{p,q}
\left(
x^p_{\scriptscriptstyle S}
+
\sum_{r}
\Bigl[
{G_{\scriptscriptstyle D}}G_{\scriptscriptstyle S}^{-1}
\Bigr]_{pr}
x^r_{\scriptscriptstyle S}
\right)
\Bigl[
\left(
I-{G_{\scriptscriptstyle D}}G_{\scriptscriptstyle S}^{-1}
\right)^{-1}
\Bigr]_{pq}
\biggl(
\sum_{r}
\Bigl[\left({G_{\scriptscriptstyle S}}-{G_{\scriptscriptstyle D}}\right)^{-1}\Bigr]_{pr}
x^r_{\scriptscriptstyle D}
$$
\begin{equation} 
\mbox{}
-
\sum_{s,r}
\Bigl[\left({G_{\scriptscriptstyle S}}-{G_{\scriptscriptstyle D}}\right)^{-1}\Bigr]_{ps}
\Bigl[
{G_{\scriptscriptstyle D}}G_{\scriptscriptstyle S}^{-1}
\Bigr]_{sr}
x^r_{\scriptscriptstyle S}
\biggr)
\label{gamma1end1}
.\end{equation} 

Note that the expressions obtained above for the parameters of the trajectory of a light signal propagating in a gravitational wave 
(\ref{gammaPend1})-(\ref{gamma1end1}) 
can be formally redefined using the further obtained delay relation for a light beam.
In this case, the numerical values of the parameters will of course not change.

For the exact model of the gravitational wave and pure radiation (\ref{Solution1})-(\ref{Solution3}), we present, omitting for brevity the calculations described above, the form of the resulting
{\bf equation of the delay of the light signal}, connecting the coordinates of the world points of the source
\(x^\alpha_{(S)}\) and the detector of the light signal \(x^\alpha_{(D)}\) in the gravitational wave in the privileged wave coordinate system:
\begin{equation}
0=
2
\left({x_{\scriptscriptstyle D}^1} - {x_{\scriptscriptstyle S}^1} \right)
+\sum_{p,q=2}^3
\left({x_{\scriptscriptstyle D}^p} - {x_{\scriptscriptstyle S}^p} \right)
\left[
\left({G_{\scriptscriptstyle D}}-{G_{\scriptscriptstyle S}}\right)^{-1}
\right]_{pq}
\left({x_{\scriptscriptstyle D}^q} - {x_{\scriptscriptstyle S}^q} \right)
\label{RetardedTimeEq}
.\end{equation}
In square brackets is the inverse matrix of the difference of the matrices
\(G_{\scriptscriptstyle D}=G^{pq}(x^0_{\scriptscriptstyle D})\) and \(G_{\scriptscriptstyle S}=G^{pq}(x^0_{\scriptscriptstyle S})\) at the world points of the detector and the source (see (\ref{functionG})). 
We~obtained the equation in ''finite differences'', i.e. in fact it is not ''local'', but integral. The obtained relation (\ref{RetardedTimeEq}) is for the gravitational wave under consideration a certain analogue of the interval along the trajectory of light propagation in flat Minkowski space-time.

The equation of the delay of radiation propagating in a gravitational wave in a simple exact analytical form (\ref{RetardedTimeEq}) has been obtained for the first time, moreover, for a non-perturbative exact model of a gravitational wave of arbitrary intensity and can be the basis for calculating various physical phenomena occurring with the participation of gravitational waves, including the calculation of the delay time of electromagnetic signals from pulsars during the passage of gravitational waves between pulsars and an observer \cite{Reardon_2023,AandArefId0,Xu_2023}.

\section{Synchronous frame of reference}

The advantage of the approach considered in the paper using the Hamilton-Jacobi formalism also includes the possibility of analytically constructing a synchronous frame of reference, which is associated with an observer freely falling in a gravitational field, with an explicitly distinguished time variable (the observer's proper time). Such a construction is based on the fact that we can analytically construct a complete integral for the action function of test particles in a privileged wave coordinate system and construct the trajectories of particle motion. By choosing a complete set of integrals of motion and the proper time of a particle for new independent variables, we obtain a transformation from a privileged wave coordinate system to a synchronous frame of reference (see \cite{LandauEng1}).

Such a synchronous frame of reference is a significant advantage for astronomical observations and allows one to analytically represent the equation for the time delay of radiation in a gravitational field, calculate the time delay of radiation in a gravitational wave, and reconstruct the characteristics of a gravitational wave from the time delay of radiation propagating against the background of a gravitational wave.

By choosing the constants \(\lambda_1\), \(\lambda_2\) and \(\lambda_3\) in the equations of the trajectory of the test particle (\ref{ParticleTraject11})-(\ref{ParticleTraject1p}) as new spatial variables, and the proper time of the particle \(\tau\) in (\ref{FunctionTau1}) as the time variable of the new reference frame, we will construct a synchronous reference frame.

For the gravitational wave metric (\ref{metric2}), the transformation from the privileged wave coordinate system  \(\{x^\alpha\}\) to the synchronous reference frame \(\{y^\alpha\}=\bigl\{t,y^1,y^2,y^3\bigr\}\) can be written from (\ref{ParticleTraject11})-(\ref{FunctionTau1}) in the following form
\begin{equation} 
x^0 \to {t} y^1
\label{TransX0}
,\end{equation} 
\begin{equation} 
x^1 \to \frac{{t}}{2 y^1}
-\frac{y^py^q}{2{\left(y^1\right)}^2}
\,
G^{pq}\bigl({t} y^1\bigr)
\label{TransX1}
,\end{equation} 
\begin{equation} 
x^p \to
\frac{y^q}{2y^1}
\,
G^{pq}\bigl({t} y^1\bigr)
,\qquad
p,q=2,3
\label{TransXp}
,\end{equation} 
\begin{equation} 
G^{pq}(x^0)=
\int{g^{pq}(x^0)}\,dx^0
,\qquad
G^{pq}\bigl({t} y^1\bigr)=G^{qp}(x^0)\Bigr\vert_{x^0 \to {t} y^1}
.\end{equation} 

The gravitational wave metric \(\tilde g{}_{\alpha\beta}\) in the synchronous reference frame \(y^\alpha\) takes the following form
\begin{equation} 
{ds}^2={dt}^2-dl^2={dt}^2+\tilde g{}_{ik}\bigl(t,y^1,y^2,y^3\bigr)\,dy^idy^k
,\qquad
i,j,k=1,2,3
,\end{equation}
where \({t}\) is the time variable, and \(y^k\) are the spatial variables of the synchronous reference frame.

For the equation of radiation delay  in a gravitational wave (\ref{RetardedTimeEq}), we can obtain in a synchronous reference system an analytical relationship between the time of signal emission at the point of the source until its detection at the point of the observer during the propagation of the signal in a gravitational wave (see f.e. \cite{OSETRIN2024169619}).

\section{Exact model of gravitational wave and pure radiation for the Bianchi type~IV universe}

As an example of using the approach proposed in the paper, we will consider the application of the results obtained above to a specific model of a gravitational wave with pure radiation in cosmological problem.

The metric for a gravitational wave (\ref{metric2}) with symmetries of the Bianchi space type IV can be represented as follows
\begin{equation} 
{ds}^2
=
2\,
{dx}^0\,{dx}^1
+
\bigl(x^0\bigr)^{(1 - \nu )}
\,
\left[
\,
\gamma^2\left(\sin\phi\right)^2\,
\left({dx^2}\right)^2
+
\Bigl(
\bigl(\log x^0-\gamma\cos{\phi}\bigr)\, {dx}^2
+
{dx^3}
\Bigr)^2
\,
\right]
\label{metric3}
,\end{equation} 
\begin{equation} 
g=\det{g_{\alpha\beta}}=-\gamma^2\sin^2\!{(\phi)}\,\bigl(x^0\bigr)^{2\left( 1 - \nu  \right) }
,\qquad
\gamma\ne 0
,\qquad
0<\phi<\pi
\label{MetricIVDown}
,\end{equation} 
where \(x^0\) is the wave variable along which the space-time interval vanishes, \(\nu\), \(\gamma\) and \(\phi\) are constant parameters of the gravitational wave model for the Bianchi type IV universe.

This model was obtained 
\cite{
OsetrinHomog2006,
Osetrin2022EPJP856} by imposing symmetries of the Bianchi space of type~{IV} on the general gravitational wave model (\ref{metric2}).
Depending on the value of the parameter \(\nu\), we obtain an expanding or collapsing model of the Bianchi type~{IV} universe with a gravitational wave.

This gravitational-wave model of space-time (\ref{metric3}) admits a three-parameter subgroup of motions, forming a homogeneity group with Killing vectors \(X_{(1)}\), \(X_{(2)}\) and \(X_{(3)}\) with a positive-definite metric on the orbits of the group:
\begin{equation} 
X^k_{(1)}=\bigl(0,0,1,0\bigr),
\qquad
X^k_{(2)}=\bigl(0,0,0,1\bigr),
\qquad
X^k_{(3)}=\bigl(-x^0,x^1,\omega x^2,\omega x^3-x^2\bigr)
,\end{equation} 
where the following notation is used
\(
\omega=(1 - \nu )/2
\).

The commutation relations for the homogeneity group Killing  vectors \(X_{(1)}\), \(X_{(2)}\) and \(X_{(3)}\) correspond to type IV according to the Bianchi classification:
\begin{equation} 
\left[X_{(1)},X_{(2)}\right]=0
,\qquad
\left[X_{(1)},X_{(3)}\right]=\omega X_{(1)}-X_{(2)}
,\qquad
\left[X_{(2)},X_{(3)}\right]=\omega X_{(2)}
.\end{equation} 

The solution (\ref{Solution2})-(\ref{Solution3}) of the Einstein equations for pure radiation 
with energy density~\(\varepsilon\) and wave vector~\(l{}^\alpha\) for the metric (\ref{metric3})  has the form:
\begin{equation} 
 l{}_\alpha=\Bigl\{  l_0, 0,0,0\Bigr\}
 ,\qquad
  l{}^\alpha=\Bigl\{ 0, l_0,0,0\Bigr\}
,\end{equation} 
\begin{equation}  \varepsilon\, l_0{}^2=\frac{{E}}{\left(x^0\right)^{2}}
,\qquad
{E}
=
\frac{-1+(1-\nu^2)\gamma^2\sin^2\!\phi}{2\gamma^2\sin^2\!\phi}
=\mbox{const}
.\end{equation} 

For cases of positive energy density of pure radiation \(\varepsilon\ge 0\), it is convenient to introduce the angular parameter \(\psi\) instead of the parameter \(\nu\):
\begin{equation}
\nu = \cos{\psi}
,\qquad
0<\psi<\pi
,\qquad
-1<\nu<1
\label{Nu}
.\end{equation}
Then we obtain
\begin{equation}
{E}=\frac{\left(\gamma\,\sin\phi\,\sin\psi\right)^2-1}{2\left(\gamma\,\sin\phi\right)^2}
,\qquad
\gamma\ne 0
,\qquad
0<\phi<\pi
,\qquad
0<\psi<\pi
\label{valueE}
.\end{equation}
%

The maximum value of the constant \({E}\) is achieved at 
values of angular 
parameters \(\phi=\psi=\pi/2\)\, (at \(\nu=0\)) and equal
\begin{equation}
{E}_{max}=\frac{\gamma^2-1}{2\,\gamma^2} = \frac{1}{2}\left(1-\frac{1}{\gamma^2}\right) ,\qquad {E}_{max}<\frac{1}{2} 
.\end{equation}

From Einstein's equations with pure radiation and the sign of the radiation energy density \(\varepsilon\) we obtain restrictions on the gravitational wave parameters 
\(\nu\) and \(\gamma\).

1.
The energy density of pure radiation \(\varepsilon\) becomes zero at the following 
values of the gravitational wave parameters:
\begin{equation} 
\vert\gamma\,\vert= \frac{1}{\sin{(\phi)}\, \sin{(\psi)}}
,\quad 
\nu = \cos{\psi}
,\quad 
0<\phi<\pi
,\quad 
0<\psi<\pi
.\end{equation}
This is the case of an expanding Bianchi IV universe.

2. The energy density of pure radiation \(\varepsilon\) is greater than zero and is given by the relations (\ref{valueE}) for the following 
values of the gravitational wave parameters:
\begin{equation} 
\vert\gamma\,\vert > \frac{1}{\sin{(\phi)}\, \sin{(\psi)}}
,\quad 
\nu = \cos{\psi}
,\quad 
0<\phi<\pi
,\quad 
0<\psi<\pi
.\end{equation} 
This is the case of an expanding Bianchi IV universe.

3.
The energy density of pure radiation \(\varepsilon\) is negative  (interpreted as radiation generation due to gravitational wave energy) for the following two ranges of parameter values:

3.A.
The value of the constant \(E\) is negative and is given by the relations (\ref{valueE}) and
\begin{equation} 
0< \vert\gamma\,\vert < \frac{1}{\sin{(\phi)}\, \sin{(\psi)}}
,\quad 
\nu = \cos{\psi}
,\quad 
0<\phi<\pi
,\quad 
0<\psi<\pi
.\end{equation} 
This is the case of an expanding Bianchi IV universe.

3.B. The value of the constant \(E\) is negative and is given by the following relations
\begin{equation} 
{E}=
\frac{1}{2}
\left(
1-\nu^2
-
\frac{1}{\gamma^2\left(\sin{\phi}\right)^2}
\right)
<0
,\qquad
\vert\nu\vert
\ge 1
,\qquad 
\gamma \ne 0
,\qquad 
0<\phi<\pi
.\end{equation} 
For \(\nu> 1\) this is the case of a collapsing Bianchi type IV universe, and for \(\nu\le 1\) this is the case of an expanding universe.

Integration of the Hamilton-Jacobi equation (\ref{HJE}) for the gravitational-wave metric (\ref{metric3}) has a special case for the parameter value \(\nu=0\), so we will consider two separate cases - when \(\nu\ne 0\) and when \(\nu=0\).


\subsection{Exact solution for gravitational wave and radiation in Bianchi type IV universe (\(\nu\ne 0\))}


In this subsection, we will use auxiliary notations: 
\begin{equation}  A=1+\gamma^2\nu^2\cos^2\!\phi>1
,\qquad
 L\bigl(y^0y^1\bigr)=1+\nu\left(\gamma\cos\phi-\log\bigl(y^0y^1\bigr)\right)
. \end{equation} 

In a privileged wave coordinate system, the equation of radiation signal delay for a gravitational wave
with metric  (\ref{metric3}) takes the following form
\[
\Delta_1 =
\frac{ -\gamma^2\nu^3\,\cos^2\!\phi }{2}
\,\biggl( A 
\left(\bigl(x_{\scriptscriptstyle D}^0\bigr)^{\nu} -\bigl(x_{\scriptscriptstyle S}^0\bigr)^{\nu}\right)^2 
-
\nu^2{\Bigl(\log (x_{\scriptscriptstyle D}^0)-
\log (x_{\scriptscriptstyle S}^0)\Bigr)}^2\bigl(x_{\scriptscriptstyle D}^0\bigr)^{\nu}\bigl(x_{\scriptscriptstyle S}^0\bigr)^{\nu}
\biggr)^{-1}
\times
\]
\[
\times
\Biggl\{
-
2\,
\left[ \,
\log (x_{\scriptscriptstyle D}^0)\,\bigl(x_{\scriptscriptstyle D}^0\bigr)^{\nu} - 
\log (x_{\scriptscriptstyle S}^0)\,\bigl(x_{\scriptscriptstyle S}^0\bigr)^{\nu} 
+\Bigl(\frac{1}{\nu} +\gamma\cos\phi \Bigr)\,
\left(\bigl(x_{\scriptscriptstyle S}^0\bigr)^{\nu}-\bigl(x_{\scriptscriptstyle D}^0\bigr)^{\nu} \right)
\right]\,\Delta_2\,\Delta_3 
\]
\[
\mbox{}
-
\biggl[
\left(
\log (x_{\scriptscriptstyle D}^0)
-2\Bigl(\frac{1}{\nu} +\gamma\cos\phi \Bigr) 
\right)
\bigl(x_{\scriptscriptstyle D}^0\bigr)^{\nu} \log (x_{\scriptscriptstyle D}^0)
-
\left(
\log (x_{\scriptscriptstyle S}^0)
-2\Bigl(\frac{1}{\nu} +\gamma\cos\phi \Bigr)
\right)
\,\bigl(x_{\scriptscriptstyle S}^0\bigr)^{\nu} \log (x_{\scriptscriptstyle S}^0)
\]
\begin{equation} 
+
\left( \frac{2
}{\nu^2} + \frac{2\,\gamma\cos\phi}{\nu}  + {\gamma }^2 \right) \,( \bigl(x_{\scriptscriptstyle D}^0\bigr)^{\nu} - \bigl(x_{\scriptscriptstyle S}^0\bigr)^{\nu})
\biggr]
\,{\Delta_2}^2  
+\left(  \bigl(x_{\scriptscriptstyle S}^0\bigr)^{\nu}-\bigl(x_{\scriptscriptstyle D}^0\bigr)^{\nu} \right)\,{\Delta_3}^2  
\Biggr\}
,\end{equation} 
where \(\Delta_k\) denotes the difference between the coordinates of  the detector and  the source
\begin{equation}  
\Delta_k=x_{\scriptscriptstyle D}^k-x_{\scriptscriptstyle S}^k
.\end{equation} 


The law of transformation of variables from a privileged wave coordinate system \(\{x^\alpha\}\) to a synchronous reference system \(\{y^\alpha\}\) takes the form:
\begin{equation} 
{{{x^0}}\rightarrow {{y^0}\,{y^1}}}
,\end{equation} 
\begin{equation} 
x^1\rightarrow 
\frac{y^0}{2y^1} - \frac{
\left(y^0y^1\right)^\nu}{2\gamma^2\nu^3\bigl(y^1\bigr)^2\cos^2\!\phi}
\left(\left(\nu y^2+L y^3\right)^2+A\bigl(y^3\bigr)^2  \right)
,\end{equation} 
\begin{equation} 
x^2\rightarrow \frac{
\left(y^0y^1\right)^\nu}{\gamma^2\nu^2y^1\cos^2\!\phi}\left(\nu y^2 + L y^3 \right)   
,\end{equation} 
\begin{equation} 
x^3\rightarrow \frac{
\left(y^0y^1\right)^\nu}{\gamma^2\nu^3y^1\cos^2\!\phi}\left(
L \left(\nu y^2 + L y^3\right)+A y^3 \right)
.\end{equation} 

The metric of the gravitational wave (\ref{metric3})  in the synchronous reference system \(\{y^\alpha\}=\bigl\{t,y^1,y^2,y^3\bigr\}\) takes the following form :
\begin{equation} {\tilde g{}_{00}} = 1
,\qquad
 {\tilde g{}_{01}} = 
{\tilde g{}_{02}} = 
{\tilde g{}_{03}} = 0
,\end{equation} 
\begin{equation}  \tilde g{}_{11} = -\frac{{{t}}^2}{\bigl(y^1\bigr)^2}+
\frac{A
\,\left({t}y^1\right)^\nu {t}}{\left(\gamma^2\cos^2\!\phi\right)^2\nu^6\bigl(y^1\bigr)^3}\left(
\left(\nu y^2+(1+L)y^3\right)^2+\gamma^2\nu^2\bigl(y^3\bigr)^2\cos^2\!\phi\right)
,\end{equation} 
\begin{equation} 
{\tilde g{}_{12}} = 
\frac{A
\,{t}\left({t}y^1\right)^\nu}{\left(\gamma^2\cos^2\!\phi\right)^2\nu^5\bigl(y^1\bigr)^2} 
\left(\nu y^2 +(1+L)y^3 \right)
,\end{equation} 
\begin{equation} 
\tilde g{}_{13} = 
\frac{A
\, {t}\left({t}y^1\right)^\nu}{\left(\gamma^2\cos^2\!\phi\right)^2\nu^6\bigl(y^1\bigr)^2}
\left( \left( 1 + L \right) \,\nu \,{y^2} + \left( {\left(1+ L \right) }^2 +\gamma^2\nu^2\cos^2\!\phi  \right)y^3 \right)
,\end{equation} 
\begin{equation} 
{\tilde g{}_{22}} = -
\frac{A\,\,{{t}}\,{\left( {{t}}\,{y^1} \right) }^{\nu }}{\left(\gamma^2\cos^2\!\phi\right)^2\,{\nu }^4\,{y^1}}
,\end{equation} 
\begin{equation} 
{\tilde g{}_{23}} = -
\frac{A\,\left(1+ L \right) \,\,{{t}}\,{\left( {{t}}\,{y^1} \right) }^{\nu }}{\left(\gamma^2\cos^2\!\phi\right)^2\,{\nu }^5\,{y^1}}
,\end{equation} 
\begin{equation} 
{\tilde g{}_{33}} = -
\frac{A
\,{{t}}\,{\left( {{t}}\,{y^1} \right) }^{\nu }}{\left(\gamma^2\cos^2\!\phi\right)^2\,{\nu }^6\,{y^1}}
\left( {\left(1+ L \right) }^2 +\gamma^2 \,{\nu }^2\cos^2\!\phi \right)
,\end{equation} 
where constants \(\gamma\), \(\nu\), \(\phi\) and \({A}\) are parameters of the gravitational wave,
variables \(y^k\) are spatial coordinates of the synchronous reference system, \(t\) is a time variable,
function \( L \) is defined by the relation
\begin{equation} 	
L(t,y^1)=1+\nu\left(\gamma\cos\phi-\log\bigl(y^1\bigr)-\log{t}\right)
\label{EqL}
.\end{equation} 

The determinant of the metric in a synchronous reference system \(\tilde g\) takes the following form:
\begin{equation} 
\tilde g=
\det{\tilde g{}_{\alpha\beta}}
=
-
\frac{{A }^2\,{{t}}^4\,{\left( {t}\,{y^1} \right) }^{2\,\nu }}{\left(\gamma^2\cos^2\!\phi \right)^3\,{\nu }^8\,\bigl(y^1\bigr)^4}
.\end{equation} 

The equation of radiation delay in a gravitational wave (\ref{RetardedTimeEq}), which relates the coordinates of the world points of the source and the detector of the signal, in a synchronous frame of reference acquires a direct physical content of the connection by separating a single time variable \({t}\). For a gravitational wave (\ref{metric3}), the equation of radiation  delay in a synchronous frame of reference can be reduced to the following form:
\[ 
\left[ 
\left(\nu{\Delta}_{12}+ {L_{\scriptscriptstyle D}}
 {\Delta}_{13}\right)^2+ A{\Delta}_{13}{}^2
\right] 
\left({t'}y_{\scriptscriptstyle S}^1\right)^\nu
-
\left[ 
\left(\nu{\Delta}_{12}+ {L_{\scriptscriptstyle S}}
 {\Delta}_{13}\right)^2+ A{\Delta}_{13}{}^2
\right] 
\left({t}y_{\scriptscriptstyle D}^1\right)^\nu
 \]
\begin{equation}  
\mbox{}
+
\gamma^2\nu^3\cos^2\!\phi
\,
\left[
\left(
\bigl({t}y_{\scriptscriptstyle D}^1\bigr)^{-\nu}-\bigl({t'}y_{\scriptscriptstyle S}^1\bigr)^{-\nu}
\right)^2
-
\frac{\nu^2}{A}\left(\log\bigl({t}y_{\scriptscriptstyle D}^1\bigr)-
\log\bigl({t'}y_{\scriptscriptstyle S}^1\bigr)\right)^2\right]y_{\scriptscriptstyle D}^1y_{\scriptscriptstyle S}^1{\Delta}_{01}=0
\label{EqDelay}
,\end{equation} 
where
\begin{equation}  
{\Delta}_{12}=y_{\scriptscriptstyle D}^2y_{\scriptscriptstyle S}^1-y_{\scriptscriptstyle D}^1y_{\scriptscriptstyle S}^2
,\ \ \ \ 
{\Delta}_{13}=y_{\scriptscriptstyle D}^3y_{\scriptscriptstyle S}^1-y_{\scriptscriptstyle D}^1y_{\scriptscriptstyle S}^3
,\ \ \ \ 
{\Delta}_{01}=y_{\scriptscriptstyle D}^1\,{t'}-y_{\scriptscriptstyle S}^1\,{t}
,\end{equation} 
\begin{equation}  
 L_{\scriptscriptstyle D}=1+\nu\left(\gamma\cos\phi-\log\bigl(y_{\scriptscriptstyle D}^1\bigr)-\log{t}\right)
 ,\quad
L_{\scriptscriptstyle S}=1+\nu\left(\gamma\cos\phi-\log\bigl(y_{\scriptscriptstyle S}^1\bigr)-\log{t'}\right)
.\end{equation} 
Here \({t'}\) is the time of signal emission by the source, \({t}\) is the time of signal detection by the observer.
The constants \(\gamma\), \(\nu\), \(\phi\) and \({A}\) are the parameters of the gravitational wave.
The spatial coordinates \(y_{\scriptscriptstyle S}^i\) specify the position of the radiation source at the moment of emission~\({t'}\), and the coordinates \(y_{\scriptscriptstyle D}^i\) specify the position of the observer detecting the signal coming from the source.

The resulting equation (\ref{EqDelay}) determines the relationship between 
the time of radiation emission \({t'}\) and the time of signal detection by 
the observer \({t}\), i.e. it gives the retarded time of the signal as it passes in a gravitational wave~(\ref{metric3}).


The equations that determine the trajectory of a test particle in a gravitational wave~(\ref{metric3}) in a synchronous frame of reference make it possible to express the spatial coordinates of the trajectory \(y^1\), \(y^2\) and \(y^3\) as functions of time \(t\).

The equation that determines  
the dependence of the spatial coordinate \(y^1\)  of 
a test particle on time variable \({t}\) can be written in the following form:
\[
  A\,\,{\left( {t}\,{y^1} \right) }^{\nu }\,
\left( 
 m^2\,{t}\,\bigl({y^1}\bigr)^2
+2\,{{\lambda }_1}\,\left( {{\lambda }_1}\,{{\sigma }_1} + 
        {{\lambda }_2}\,{{\sigma }_2} + {{\lambda }_3}\,{{\sigma }_3} \right) \,
      {y^1} 
-{{{\lambda }_1}}^2\,{t} 
\right)
+\mbox{}
\]
\begin{equation}      
\mbox{}
+
{y^1}
\gamma^2\cos^2\!\phi\,\nu \,{{{\lambda }_1}}^2\,
   \left( A\,{{{\sigma }_2}}^2 
+ \left[ {{\sigma }_2} -  \nu \,{{\sigma }_3}+\nu{{\sigma }_2}\,\left(\gamma\cos\phi-\log y^1-\log{t}\right)
 \right]^2 
\right) 
=0
\label{y1Eq}
.\end{equation} 

The equation (\ref{y1Eq}) defines, albeit implicitly, the spatial coordinate of the trajectory of a test particle \(y^1\) as a function of time \(t\). Then the remaining spatial coordinates of the particle trajectory \(y^2\) and \(y^3\) are determined through \(y^1(t)\) by the following trajectory equations:

\begin{equation} 	
{{{y^2}}(t)=
    {{y^1}\,
\Biggl(
\frac{{{\lambda }_2}}{{{\lambda }_1}} 
+
\frac{
\gamma^2 \,\nu \,\cos^2\!\phi\,
\bigl(
\left( A + L^2 \right) \,{{\sigma }_2} -  L\,\nu \,{{\sigma }_3} 
\bigr)
}{A\,{\left( {t}\,{y^1} \right) }^{\nu }} 
\Biggr) }}
,\end{equation} 
\begin{equation} 
  {{{y^3}}(t)=
    {{y^1}\,
\Biggl(
\frac{{{\lambda }_3}}{{{\lambda }_1}} -
        \frac{\gamma^2\,{\nu }^2 \cos^2\!\phi \,
           \left( L\,{{\sigma }_2} - \nu \,{{\sigma }_3} \right) }
           {A\,{\left( {t}\,{y^1} \right) }^{\nu }} 
\Biggr) 
}}
.\end{equation} 
where \(\lambda_i\) and \(\sigma_i\) are constant parameters of the trajectory of test particle, \(\gamma\), \(\nu\), \(\phi\), \({A}\) are constant parameters of the gravitational wave, and the function \(L\) is determined by the relation (\ref{EqL}).

The trajectories of a light beam in a gravitational wave in a synchronous frame of reference can be written in such a way as to single out the equation that relates the spatial coordinate of the trajectory \(y^1\) to the time variable \(t\). Then the spatial coordinates of the light trajectory \(y^2\) and \(y^3\) will be determined through \(y^1(t)\):
\[
  A\,\,{\left( {t}\,{y^1} \right) }^{\nu }\,
   \left( {{{\kappa}_1}}\,{t} - 
     2\,\left( {{\kappa}_1}\,{{\gamma}_1} + 
        {{\kappa}_2}\,{{\gamma}_2} + {{\kappa}_3}\,{{\gamma}_3} \right) \,
      {y^1} \right)
\]
\begin{equation}  
\mbox{}
-     
\gamma^2\,\nu \,{{{\kappa}_1}}\,\cos^2\!\phi\,
   \left( A\,{{{\gamma}_2}}^2 + {\left( L\,{{\gamma}_2} - 
         \nu \,{{\gamma}_3} \right) }^2 \right) \,{y^1}=0
,\end{equation} 
\begin{equation} 	
{{y^2}(t)=
    {{y^1}\,\left( \frac{{{\kappa}_2}}{{{\kappa}_1}} +
        \frac{\gamma^2\cos^2\!\phi \,\nu \,
           \left( \left( A + L^2 \right) \,{{\gamma}_2} - 
             L\,\nu \,{{\gamma}_3} \right) }{A\,
           \,{\left( {t}\,{y^1} \right) }^{\nu }} \right) }}
,\end{equation} 
\begin{equation} 
  {{y^3}(t)=
    {{y^1}\,\left( \frac{{{\kappa}_3}}{{{\kappa}_1}} - 
        \frac{\gamma^2\cos^2\!\phi \,{\nu }^2\,
           \left( L\,{{\gamma}_2} - \nu \,{{\gamma}_3} \right) }
           {A\,\,{\left( {t}\,{y^1} \right) }^{\nu }} \right) }}
,\end{equation} 
where \(k_i\) and \(\gamma_i\) are constant parameters of the trajectory of light beam, \(\gamma\), \(\nu\), \(\phi\), \({A}\) are constant parameters of the gravitational wave, and the function \(L\) is determined by the relation (\ref{EqL}).

\subsection{Exact solution for gravitational wave and radiation in Bianchi type IV universe (\(\nu= 0\))}

The peculiarity that arises when integrating the Hamilton-Jacobi equation of test particles for the gravitational wave metric in a Bianchi type IV universe when the parameter~\(\nu\) becomes zero requires that this case be considered separately.

The metric of a gravitational wave in a privileged wave coordinate system  in this case  takes on the following special form:
\begin{equation} 
{ds}^2
=
2\,
{dx}^0\,{dx}^1
+
x^0
\,
\left[
\,
\gamma^2\left(\sin\phi\right)^2\,
\left({dx^2}\right)^2
+
\Bigl(
\bigl(\log x^0-\gamma\cos{\phi}\bigr)\, {dx}^2
+
{dx^3}
\Bigr)^2
\,
\right]
\label{metric4}
,\end{equation} 
where \(x^0\) is the wave variable,
constants \(\gamma\) and \(\phi\) (\(0<\phi<\pi\)) are the parameters of the gravitational wave.

The solution (\ref{Solution2})-(\ref{Solution3}) of the Einstein equations for pure radiation 
with energy density~\(\varepsilon\) and wave vector~\(l{}^\alpha\) for the metric (\ref{metric4})  has the form:
\begin{equation} 
 l{}_\alpha=\Bigl\{  l_0, 0,0,0\Bigr\}
 ,\qquad
  l{}^\alpha=\Bigl\{ 0, l_0,0,0\Bigr\}
,\end{equation} 
\begin{equation}  \varepsilon\, l_0{}^2=\frac{{E}}{\left(x^0\right)^{2}}
,\qquad
{E}
=
\frac{{\gamma^2}-\csc^2({\phi})}{2 {\gamma^2}}
=\mbox{const}
.\end{equation} 

The energy density of pure radiation \(\varepsilon\) has a positive value for the following range of parameter values:
\begin{equation} 
\vert\gamma\vert>\frac{1}{\sin(\phi)}
,\qquad
0<\phi<\pi
.\end{equation} 

The result of integrating the Hamilton-Jacobi equation for test particles in the privileged wave coordinate system (\ref{ParticleTraject11})-(\ref{FunctionTau1}) for the considered case of the gravitational wave will take the following form:
\begin{equation} 
x^0({\tau}) = {\lambda_1} \bigl( {\tau} - {\tau}_0\bigr)
\label{TrajectoryNu0x0}
,\end{equation} 
$$
x^1({\tau}) =
\sigma_1
 -\frac{
 1 }{4 {\gamma}^2 {\lambda_1}^2 \sin^2({\phi}) } 
 \,
\biggl[
 2 \log ({\lambda_1} {\tau}) 
\Bigl(
 {\gamma}^2 {\lambda_3}^2+2 {\gamma} {\lambda_2} {\lambda_3} \cos ({\phi})+{\lambda_2}^2
\Bigr)
$$
\begin{equation} 
\mbox{}
 -2 {\gamma}^2 {\lambda_1} {\tau} \sin ^2({\phi})-2 {\lambda_3} \log ^2({\lambda_1} {\tau}) ({\gamma} {\lambda_3} \cos ({\phi})+{\lambda_2})+\frac{2}{3} {\lambda_3}^2 \log ^3({\lambda_1} {\tau})
\biggr]
\label{TrajectoryNu0x1}
,\end{equation} 
\begin{equation} 
x^2({\tau}) = 
\sigma_2
+
\frac{
\log ({\lambda_1} {\tau})
}{2 {\gamma}^2 {\lambda_1}\sin^2({\phi}) } 
\Bigl(
2 {\lambda_2}
+
2 {\gamma} {\lambda_3} \cos ({\phi})-{\lambda_3} \log ({\lambda_1} {\tau})
\Bigr)
\label{TrajectoryNu0x2}
,\end{equation} 
\[
x^3({\tau}) = 
\sigma_3
+
\frac{
\log ({\lambda_1} {\tau}) 
}{6 {\gamma}^2 {\lambda_1}\sin^2({\phi}) } 
\Bigl(
6 {\gamma}^2 {\lambda_3}+6 {\gamma} \cos ({\phi}) ({\lambda_2}-{\lambda_3} \log ({\lambda_1} {\tau}))
\]
\begin{equation} 
-3 {\lambda_2} \log ({\lambda_1} {\tau})+2 {\lambda_3} \log ^2({\lambda_1} {\tau})
\Bigr)
\label{TrajectoryNu0x3}
,\end{equation} 
where \(\tau\) is the proper time of the particle,  constants \(\lambda_k\), \(\sigma_k\) and \(\tau_0\)  are parameters determined by the initial or boundary conditions of the particle motion.

The delay time equation for the propagation of radiation in a gravitational wave (\ref{RetardedTimeEq}) for the case under consideration  in the privileged wave coordinate system will take the following form:
\[
0=\Delta_1
+
\frac{6 {\gamma}^2  \sin ^2({\phi}) 
\Bigl(
  \log ({x^0_{\scriptscriptstyle D}}{x^0_{\scriptscriptstyle S}})
-2 {\gamma}  \cos ({\phi})
\Bigr)
\,
{\Delta_2}{\Delta_3}
}{
\log ({x^0_{\scriptscriptstyle D}}/{x^0_{\scriptscriptstyle S}}) 
\,
 \left(
 6 {\gamma}^2
 -6 {\gamma}^2 \cos (2 {\phi})
 +\log^2 ({x^0_{\scriptscriptstyle D}}/{x^0_{\scriptscriptstyle S}})
 \right)
}
\]
\[
\mbox{}
+
\frac{2 {\gamma}^2  \sin ^2({\phi}) 
\Bigl(
3 {\gamma}^2-3 {\gamma} \cos ({\phi}) 
\log ({x^0_{\scriptscriptstyle D}}{x^0_{\scriptscriptstyle S}})
+\log ({x^0_{\scriptscriptstyle D}}) \log
   ({x^0_{\scriptscriptstyle S}})+\log ^2({x^0_{\scriptscriptstyle D}})+\log ^2({x^0_{\scriptscriptstyle S}})\Bigr)
 \,{\Delta_2}^2
}{
\log ({x^0_{\scriptscriptstyle D}}/{x^0_{\scriptscriptstyle S}}) 
\,
 \left(
 6 {\gamma}^2
 -6 {\gamma}^2 \cos (2 {\phi})
 +\log^2 ({x^0_{\scriptscriptstyle D}}/{x^0_{\scriptscriptstyle S}})
 \right)
}
\]
\begin{equation} 
\mbox{}
+
\frac{6 {\gamma}^2   \sin ^2({\phi})
\,{\Delta_3}^2 
}{
\log ({x^0_{\scriptscriptstyle D}}/{x^0_{\scriptscriptstyle S}}) 
\,
 \left(
 6 {\gamma}^2
 -6 {\gamma}^2 \cos (2 {\phi})
 +\log^2 ({x^0_{\scriptscriptstyle D}}/{x^0_{\scriptscriptstyle S}})
 \right)
}
,\end{equation} 
where
\(x_{\scriptscriptstyle D}^\alpha\) are the coordinates of the world point of the detector of signal,
\(x_{\scriptscriptstyle S}^\alpha\) are the coordinates of the world point of the signal source, 
\(x^0\) is the wave variable,
constants \(\gamma\) and \(\phi\) (\(0<\phi<\pi\)) are the parameters of the gravitational wave,
and
\[
\Delta_k={x_{\scriptscriptstyle D}^k} - {x_{\scriptscriptstyle S}^k}
,\qquad
i,j,k=1,2,3
.\]

In accordance with the solutions of the equations of motion of test particles in the Hamilton-Jacobi formalism 
  for this case (\ref{TrajectoryNu0x0})-(\ref{TrajectoryNu0x3}), 
we obtain the following formulas for the transition from the privileged wave coordinate system \(\{x^\alpha\}\) to the synchronous reference system \(\{y^\alpha\}=\left(\tau,y^1,y^2,y^3 \right)\):
$$
x^0  \to  {y^1} {\tau} 
, $$
$$
x^1  \to 
-\frac{
1}{4 {\gamma}^2 {y^1}^2\sin^2({\phi}) } 
\,
\biggl(
2 \log ({y^1} {\tau}) \left({\gamma}^2 {y^3}^2+2 {\gamma} {y^2} {y^3} \cos ({\phi})+{y^2}^2\right)
-2 {\gamma}^2 {y^1} {\tau} \sin ^2({\phi})
$$
\begin{equation} 
-2 {y^3} \log ^2({y^1} {\tau}) ({\gamma} {y^3} \cos ({\phi})
+{y^2})+\frac{2}{3} {y^3}^2 \log ^3({y^1} {\tau})
\biggr)
,\end{equation} 
\begin{equation} 
x^2  \to  \frac{
\log ({y^1} {\tau}) \left(2 {\gamma} {y^3} \cos ({\phi})-{y^3} \log ({y^1} {\tau})+2 {y^2}\right)
}{2 {\gamma}^2 {y^1}\sin^2({\phi}) } 
,\end{equation} 
\[
x^3  \to  
\frac{
\log ({y^1} {\tau}) 
}{6 {\gamma}^2 {y^1}\sin^2({\phi}) } 
\,
\biggl(
6 {\gamma}^2 {y^3}+6 {\gamma} \cos ({\phi}) \left({y^2}-{y^3} \log ({y^1} {\tau})\right)
\]
\begin{equation} 
\mbox{}
-3 {y^2} \log ({y^1} {\tau})+2 {y^3} \log ^2({y^1} {\tau})
\biggr)
.\end{equation} 
Here \({\tau}\) is a time variable, and the variables \({y^k}\) are spatial coordinates.

The components of the metric of the gravitational wave (\ref{metric4})  in the synchronous reference system will take the following form:
\[
\tilde g{}^{00} = 1 
,\quad
\tilde g{}^{01} = 0 
,\quad
\tilde g{}^{02} = 0 
,\quad
\tilde g{}^{03} = 0 
,\]
\[
\tilde g{}^{11} = -\frac{{y^1}^2}{{\tau}^2} 
,\qquad
\tilde g{}^{12} = -\frac{{y^1} {y^2}}{{\tau}^2} 
,\qquad
\tilde g{}^{13} = -\frac{{y^1} {y^3}}{{\tau}^2} 
,\]
$$
\tilde g{}^{22} =
-\frac{{y^2}^2}{{\tau}^2} 
+
 \frac{4 {\gamma}^2 {y^1} \sin ^2({\phi}) 
 }{{\tau} \log ^2({y^1} {\tau}) 
 \left(
6 {\gamma}^2\bigl(1-\cos (2 {\phi})\bigr)
+\log ^2({y^1} {\tau})\right)^2}
\,
\Bigl(
3 {\gamma}^2 \log ^2({y^1} {\tau})
$$
\[
\mbox{}
+18 {\gamma}^4
-6 {\gamma}^2 \cos (2 {\phi}) \left(3 {\gamma}^2-\log ^2({y^1} {\tau})\right)
-6 {\gamma} \cos ({\phi}) \log ^3({y^1} {\tau})
+\log ^4({y^1} {\tau})
\Bigr)
,\]
\[
\tilde g{}^{23} = 
-\frac{{y^2} {y^3}}{{\tau}^2} 
+
\frac{
12 {\gamma}^2 {y^1}\! \sin ^2({\phi}) 
\Bigl(
3 {\gamma}^3\! \cos (3 {\phi})
-3 {\gamma} \cos ({\phi}) 
\bigl({\gamma}^2+\log ^2({y^1} {\tau})\bigr)
+\log^3({y^1} {\tau})
\Bigr)
}{{\tau} \log ^2({y^1} {\tau}) \left(
6 {\gamma}^2
\bigl(
1-\cos (2 {\phi})
\bigr)
+\log ^2({y^1} {\tau})\right)^2}
,\]
\[
\tilde g{}^{33} =
 -\frac{{y^3}^2}{{\tau}^2} 
+
 \frac{
 36 {\gamma}^2 {y^1} \sin ^2({\phi}) 
\Bigl(
\log ^2({y^1} {\tau})
 -
 2 {\gamma}^2 \cos (2 {\phi})
+
 2 {\gamma}^2
\Bigr)
 }{{\tau} \log ^2({y^1} {\tau}) \left(
 6 {\gamma}^2
\bigl(
1-\cos (2 {\phi})
\bigr)
 +\log ^2({y^1} {\tau})\right)^2}
,\] 
where \(\tau\) is the time variable, the variables \({y^k}\) are spatial coordinates,
constants \(\gamma\) and \(\phi\) (\(0<\phi<\pi\)) are the parameters of the gravitational wave.

The retarded time equation for the propagation of radiation (\ref{RetardedTimeEq}) in a gravitational wave (\ref{metric4}) can be written in a synchronous frame of reference with a time variable \(\tau\) in the following form:
\[
0=
\csc ({\phi}) \log ^3({\tau'} y_{\scriptscriptstyle S}^{1}) 
\biggl[
6 {\gamma}{}^2 y_{\scriptscriptstyle D}^{1} y_{\scriptscriptstyle S}^{1} \sin ^2({\phi}) ({\tau} y_{\scriptscriptstyle S}^{1}-{\tau'}
   y_{\scriptscriptstyle D}^{1})
\]
\[
\mbox{}
-6 \log ({\tau} y_{\scriptscriptstyle D}^{1}) \left({\gamma}{}^2 (y_{\scriptscriptstyle D}^{3} y_{\scriptscriptstyle S}^{1}-y_{\scriptscriptstyle D}^{1} y_{\scriptscriptstyle S}^{3})^2+(y_{\scriptscriptstyle D}^{1})^2 (y_{\scriptscriptstyle S}^{2})^2-2 y_{\scriptscriptstyle D}^{1} y_{\scriptscriptstyle D}^{2} y_{\scriptscriptstyle S}^{1}   y_{\scriptscriptstyle S}^{2}+(y_{\scriptscriptstyle D}^{2})^2 (y_{\scriptscriptstyle S}^{1})^2\right)
\]
\[
\mbox{}
+2 \log ({\tau} y_{\scriptscriptstyle D}^{1}) (y_{\scriptscriptstyle D}^{3} y_{\scriptscriptstyle S}^{1}-y_{\scriptscriptstyle D}^{1} y_{\scriptscriptstyle S}^{3}) 
\biggl(
3 {\gamma} \cos ({\phi}) 
\Bigl(
\log ({\tau} y_{\scriptscriptstyle D}^{1})
(y_{\scriptscriptstyle D}^{3} y_{\scriptscriptstyle S}^{1}-y_{\scriptscriptstyle D}^{1} y_{\scriptscriptstyle S}^{3})
+2 y_{\scriptscriptstyle D}^{1} y_{\scriptscriptstyle S}^{2}-2 y_{\scriptscriptstyle D}^{2} y_{\scriptscriptstyle S}^{1}
\Bigr)
\]
\[
\mbox{}
+\log ({\tau} y_{\scriptscriptstyle D}^{1}) 
\Bigl(
\log ({\tau} y_{\scriptscriptstyle D}^{1}) 
(y_{\scriptscriptstyle D}^{1} y_{\scriptscriptstyle S}^{3}-y_{\scriptscriptstyle D}^{3}   y_{\scriptscriptstyle S}^{1})
-3 y_{\scriptscriptstyle D}^{1} y_{\scriptscriptstyle S}^{2}+3 y_{\scriptscriptstyle D}^{2} y_{\scriptscriptstyle S}^{1}
\Bigr)
\biggr)
\biggr]
\]
\[
\mbox{}
+3 \csc ({\phi}) \log ({\tau} y_{\scriptscriptstyle D}^{1}) \log ^2({\tau'} y_{\scriptscriptstyle S}^{1}) 
\biggl[
6 {\gamma}{}^2 y_{\scriptscriptstyle D}^{1} y_{\scriptscriptstyle S}^{1} \sin ^2({\phi}) ({\tau'} y_{\scriptscriptstyle D}^{1}-{\tau} y_{\scriptscriptstyle S}^{1})
\]
\[
\mbox{}
+2 \log ^2({\tau} y_{\scriptscriptstyle D}^{1}) (y_{\scriptscriptstyle D}^{3} y_{\scriptscriptstyle S}^{1}-y_{\scriptscriptstyle D}^{1} y_{\scriptscriptstyle S}^{3}) 
\Bigl(
{\gamma} \cos   ({\phi}) (y_{\scriptscriptstyle D}^{3} y_{\scriptscriptstyle S}^{1}-y_{\scriptscriptstyle D}^{1} y_{\scriptscriptstyle S}^{3})-y_{\scriptscriptstyle D}^{1} y_{\scriptscriptstyle S}^{2}+y_{\scriptscriptstyle D}^{2} y_{\scriptscriptstyle S}^{1}
\Bigr)
\]
\[
\mbox{}
+2 \log ({\tau} y_{\scriptscriptstyle D}^{1}) 
\biggl(
-2 {\gamma} (y_{\scriptscriptstyle D}^{3} y_{\scriptscriptstyle S}^{1}-y_{\scriptscriptstyle D}^{1} y_{\scriptscriptstyle S}^{3}) 
\Bigl(
{\gamma} \cos (2 {\phi}) 
(y_{\scriptscriptstyle D}^{3} y_{\scriptscriptstyle S}^{1}-y_{\scriptscriptstyle D}^{1} y_{\scriptscriptstyle S}^{3}) 
+\cos ({\phi}) (y_{\scriptscriptstyle D}^{2} y_{\scriptscriptstyle S}^{1}-y_{\scriptscriptstyle D}^{1}   y_{\scriptscriptstyle S}^{2})
\Bigr)
\]
\[
\mbox{}
+({\gamma} y_{\scriptscriptstyle D}^{1} y_{\scriptscriptstyle S}^{3}-{\gamma} y_{\scriptscriptstyle D}^{3} y_{\scriptscriptstyle S}^{1}+y_{\scriptscriptstyle D}^{1} y_{\scriptscriptstyle S}^{2}-y_{\scriptscriptstyle D}^{2} y_{\scriptscriptstyle S}^{1}) ({\gamma} y_{\scriptscriptstyle D}^{1}
   y_{\scriptscriptstyle S}^{3}-{\gamma} y_{\scriptscriptstyle D}^{3} y_{\scriptscriptstyle S}^{1}-y_{\scriptscriptstyle D}^{1} y_{\scriptscriptstyle S}^{2}+y_{\scriptscriptstyle D}^{2} y_{\scriptscriptstyle S}^{1})
\biggr)
\biggr]
\]
\[
\mbox{}
+6 {\gamma}{}^2 y_{\scriptscriptstyle D}^{1} y_{\scriptscriptstyle S}^{1} \log ({\tau} y_{\scriptscriptstyle D}^{1}) ({\tau'}
   y_{\scriptscriptstyle D}^{1}-{\tau} y_{\scriptscriptstyle S}^{1}) \left(\sin ({\phi}) \left(9 {\gamma}{}^2+\log ^2({\tau} y_{\scriptscriptstyle D}^{1})\right)-3 {\gamma}{}^2 \sin (3 {\phi})\right)
\]
\[
\mbox{}
-3 \csc ({\phi}) \log ({\tau'} y_{\scriptscriptstyle S}^{1}) 
\biggl[
24 {\gamma}{}^4 y_{\scriptscriptstyle D}^{1} y_{\scriptscriptstyle S}^{1} \sin ^4({\phi}) ({\tau'} y_{\scriptscriptstyle D}^{1}-{\tau} y_{\scriptscriptstyle S}^{1})
\]
\[
\mbox{}
+6 {\gamma}{}^2 \sin^2({\phi}) \log ({\tau} y_{\scriptscriptstyle D}^{1}) 
\biggl(
y_{\scriptscriptstyle D}^{1} y_{\scriptscriptstyle S}^{1} \log ({\tau} y_{\scriptscriptstyle D}^{1}) ({\tau'} y_{\scriptscriptstyle D}^{1}-{\tau} y_{\scriptscriptstyle S}^{1})
\]
\[
\mbox{}
+
4\, \left(
{\gamma}{}^2 (y_{\scriptscriptstyle D}^{3} y_{\scriptscriptstyle S}^{1}-y_{\scriptscriptstyle D}^{1} y_{\scriptscriptstyle S}^{3})^2+(y_{\scriptscriptstyle D}^{1})^2 (y_{\scriptscriptstyle S}^{2})^2
-2 y_{\scriptscriptstyle D}^{1} y_{\scriptscriptstyle D}^{2} y_{\scriptscriptstyle S}^{1} y_{\scriptscriptstyle S}^{2}+(y_{\scriptscriptstyle D}^{2})^2 (y_{\scriptscriptstyle S}^{1})^2
\right)
\biggr)
\]
\[
\mbox{}
+4 {\gamma} \cos ({\phi}) \log
   ({\tau} y_{\scriptscriptstyle D}^{1}) (y_{\scriptscriptstyle D}^{2} y_{\scriptscriptstyle S}^{1}-y_{\scriptscriptstyle D}^{1} y_{\scriptscriptstyle S}^{2}) (y_{\scriptscriptstyle D}^{3} y_{\scriptscriptstyle S}^{1}-y_{\scriptscriptstyle D}^{1} y_{\scriptscriptstyle S}^{3}) \left(12 {\gamma}{}^2 \sin ^2({\phi})+\log ^2({\tau}
   y_{\scriptscriptstyle D}^{1})\right)
\]
\begin{equation} 
\mbox{}
   +2 \log ^3({\tau} y_{\scriptscriptstyle D}^{1}) \left({\gamma}{}^2 (y_{\scriptscriptstyle D}^{3} y_{\scriptscriptstyle S}^{1}-y_{\scriptscriptstyle D}^{1} y_{\scriptscriptstyle S}^{3})^2+(y_{\scriptscriptstyle D}^{1})^2 (y_{\scriptscriptstyle S}^{2})^2-2 y_{\scriptscriptstyle D}^{1} y_{\scriptscriptstyle D}^{2} y_{\scriptscriptstyle S}^{1}
   y_{\scriptscriptstyle S}^{2}+(y_{\scriptscriptstyle D}^{2})^2 (y_{\scriptscriptstyle S}^{1})^2
\right)
\biggr]
\label{RetardedTimeEq4}
.\end{equation} 
The retarded time equation for the propagation of radiation (\ref{RetardedTimeEq4})
gives us the relationship between the signal detection time \(\tau\) and the emission time of this signal \(\tau'\) when the signal passes in a gravitational wave.
The resulting equation contains the spatial coordinates of the source \(y_{\scriptscriptstyle S}^k\) and the spatial coordinates of the radiation detector \(y_{\scriptscriptstyle D}^k\) (observer).



\section{Conclusion}

The paper considers a general exact model of a gravitational wave, with a space-time metric depending in a privileged coordinate system on one wave variable and with sources in the form of pure radiation, dust matter and a cosmological constant. It is shown that the compatibility conditions of Einstein's field equations for this model lead to a contradiction with the presence of dust matter in the model. An exact solution of the field equations for a gravitational wave with pure radiation is obtained. Light trajectories, radiation propagation characteristics are found.
The retarded time equation 
of light signals in a gravitational wave is obtained.
Based on the obtained general results, an exact model of gravitational wave and pure radiation for the Bianchi universe of type IV is constructed.

\authorcontributions{Conceptualization, K.O.; methodology, K.O. and V.E.; validation, V.E., A.F. and K.O.;  investigation, K.O., V.E. and A.F.; writing---original draft preparation, K.O.; supervision, K.O.; project administration, V.E.; funding acquisition, V.E. All authors have read and agreed to the published version of the manuscript.}

\funding{The study was supported by the Russian Science Foundation, grant \mbox{No. 23-22-00343},
\url{https://rscf.ru/en/project/23-22-00343/}}

\conflictsofinterest{The authors declare no conflict of interest. }

%

\dataavailability{Data are contained within the article.} 


%

\appendixtitles{yes} 
\appendixstart
\appendix
\section[\appendixname~\thesection]{Parameters of light ray trajectories in exact model of gravitational wave}
\label{AppendixA}

This appendix presents a variant of calculating the parameters for the trajectory of radiation propagation (\ref{LightTraject11})-(\ref{LightTraject1p})  against the background of a gravitational wave (\ref{metric2}) between the world point of the radiation source and the world point of signal detection by the observer.

Let us denote the variables of the radiation source (light signal) as $x^\alpha_{\scriptscriptstyle S}$, and the variables of the observer (detector) as $x^\alpha_{\scriptscriptstyle D}$, then we obtain a system of equations for the trajectory of the light beam connecting these points (points $x^\alpha_{\scriptscriptstyle S}$ and $x^\alpha_{\scriptscriptstyle D}$):
\begin{equation} 
x^1_{\scriptscriptstyle S}=\gamma_1-\frac{k_pk_q}{2{\left(k_1\right)}^2}
\,
G^{pq}\bigl(x^0_{\scriptscriptstyle S}\bigr)
\label{LightEq1S}
,\end{equation} 
\begin{equation} 
x^p_{\scriptscriptstyle S}=\gamma_p+\frac{k_q}{k_1}
\,
G^{pq}\bigl(x^0_{\scriptscriptstyle S}\bigr)
\label{LightEqpS}
,\end{equation} 
\begin{equation} 
x^1_{\scriptscriptstyle D}=\gamma_1-\frac{k_pk_q}{2{\left(k_1\right)}^2}
\,
G^{pq}\bigl(x^0_{\scriptscriptstyle D}\bigr)
\label{LightEq1D}
,\end{equation} 
\begin{equation} 
x^p_{\scriptscriptstyle D}=\gamma_p+\frac{k_q}{k_1}
\,
G^{pq}\bigl(x^0_{\scriptscriptstyle D}\bigr)
\label{LightEqpD}
,\end{equation} 
where
\[
G^{pq}(x^0)=
\int{g^{pq}(x^0)}\,dx^0
,\qquad
G_{\scriptscriptstyle D}=G^{pq}(x^0_{\scriptscriptstyle D})
,\qquad
G_{\scriptscriptstyle S}=G^{pq}(x^0_{\scriptscriptstyle S})
.\]

To shorten the text, we will use the notation $G_{\scriptscriptstyle S}$ for the matrix $G^{pq}\bigl(x^0_{\scriptscriptstyle S}\bigr)$ and the notation $G_{\scriptscriptstyle D}$ for the matrix $G^{pq}\bigl(x^0_{\scriptscriptstyle D}\bigr)$:
\begin{equation} 
G^{pq}\bigl(x^0_{\scriptscriptstyle S}\bigr)=\left[G_{\scriptscriptstyle S}\right]^{pq}
,\quad
G^{pq}\bigl(x^0_{\scriptscriptstyle D}\bigr)=\left[G_{\scriptscriptstyle D}\right]^{pq}
,\end{equation} 
where $G(x^0)$ is a matrix constructed from the integrals of the metric components.

To obtain the ''signal delay'' relation, i.e. the relation linking the variables
$x^\alpha_{\scriptscriptstyle D}$ and $x^\alpha_{\scriptscriptstyle S}$, it is necessary to exclude the light signal parameters $\gamma_1$, $\gamma_2$, $\gamma_3$, $k_2/k_1$ and $k_3/k_1$ from the equations
(\ref{LightEq1S})-(\ref{LightEqpD}).
As a result, there remains one relation linking the coordinates of the world points of the source and detector $x^\alpha_{\scriptscriptstyle S}$ and $x^\alpha_{\scriptscriptstyle D}$.

Thus, from the equations (\ref{LightEqpS}) we obtain an expression for $k_p/k_1$ through the parameters $\gamma_p$:
\begin{equation} 
\frac{k_p}{k_1}=\sum_{q}\left[G_{\scriptscriptstyle S}^{-1}\right]_{pq}\left(x^q_{\scriptscriptstyle S}-\gamma_q\right)
\label{kappaPfirst}
.\end{equation} 
Then from equation (\ref{LightEq1S}) taking into account (\ref{kappaPfirst}) we obtain the relation for $\gamma_1$ through the parameters $\gamma_p$:
\begin{equation} 
\gamma_1=
x^1_{\scriptscriptstyle S}+\frac{k_pk_q}{2{\left(k_1\right)}^2}
\,
G^{pq}\bigl(x^0_{\scriptscriptstyle S}\bigr)
=
x^1_{\scriptscriptstyle S}+
\frac{1}{2}
\sum_{p,q}
\left[G_{\scriptscriptstyle S}^{-1}\right]_{pq}
\left(x^p_{\scriptscriptstyle S}-\gamma_p\right)
\left(x^q_{\scriptscriptstyle S}-\gamma_q\right)
\label{gamma1first}
.\end{equation} 
Equation (\ref{LightEqpD}) taking into account (\ref{kappaPfirst}) can be written as follows
$$
x^p_{\scriptscriptstyle D}=\gamma_p+\frac{k_q}{k_1}
\,
G^{pq}\bigl(x^0_{\scriptscriptstyle D}\bigr)=
\gamma_p+
\sum_{r,q}
G^{pq}_{\scriptscriptstyle D}
\left[G_{\scriptscriptstyle S}^{-1}\right]_{qr}\left(x^r_{\scriptscriptstyle S}-\gamma_r\right)
=
$$
\begin{equation} 
=
\sum_r
\gamma_r
\left(
\delta^r_p-
\sum_{q}
G^{pq}_{\scriptscriptstyle D}
\left[G_{\scriptscriptstyle S}^{-1}\right]_{qr}
\right)
+
\sum_{r,q}
G^{pq}_{\scriptscriptstyle D}
\left[G_{\scriptscriptstyle S}^{-1}\right]_{qr}x^r_{\scriptscriptstyle S}
\label{EqGammaP}
.\end{equation} 
The obtained relation (\ref{EqGammaP}) allows us to determine the parameters $\gamma_p$ for the light signal through the coordinates of the source and the detector, using matrix notations to shorten the notation
\begin{equation} 
\gamma_p=
\sum_q
\Bigl[
\left(
I-{G_{\scriptscriptstyle D}}G_{\scriptscriptstyle S}^{-1}
\right)^{-1}
\Bigr]_{pq}
\left(
x^q_{\scriptscriptstyle D}
-
\sum_{r}
\Bigl[
{G_{\scriptscriptstyle D}}G_{\scriptscriptstyle S}^{-1}
\Bigr]_{qr}
x^r_{\scriptscriptstyle S}
\right)
\label{gammaPend}
,\end{equation} 
where $G(x^0)$ is the matrix of integrals of the metric components in the relations (\ref{functionG}), $I$ is the identity matrix, for the designation of whose components the Kronecker symbols $\delta^q_p$ are also used.

Using the obtained relation (\ref{gammaPend}) for $\gamma_p$
we can find from the relation (\ref{gamma1first}) the required form of the parameter $\gamma_1$ for the light signal from the world point $x^\alpha_{\scriptscriptstyle S}$ of the source to the world point $x^\alpha_{\scriptscriptstyle D}$ of the detector. For $\gamma_1$ we have:
$$
\gamma_1=
x^1_{\scriptscriptstyle S}
+
\frac{1}{2}
\sum_{p,q}
x^p_{\scriptscriptstyle S}
\Bigl(
\left[G_{\scriptscriptstyle S}^{-1}\right]_{pq} x^q_{\scriptscriptstyle S}-
\left[G_{\scriptscriptstyle S}^{-1}\right]_{pq} \gamma_q
\Bigr)
$$
$$
\mbox{}
-
\frac{1}{2}
\sum_{r,p}
%
\left(
x^p_{\scriptscriptstyle S}
-
\sum_{q}
\Bigl[
{G_{\scriptscriptstyle D}}G_{\scriptscriptstyle S}^{-1}
\Bigr]_{pq}
x^q_{\scriptscriptstyle S}
\right)
\left(
\Bigl[
\left(
{G_{\scriptscriptstyle S}}-{G_{\scriptscriptstyle D}}
\right)^{-1}\Bigr]_{rp}x^r_{\scriptscriptstyle S}
-
\Bigl[
\left(
{G_{\scriptscriptstyle S}}-{G_{\scriptscriptstyle D}}
\right)^{-1}\Bigr]_{rp}\gamma_r
\right)
=
$$
$$
=
x^1_{\scriptscriptstyle S}
+
\frac{1}{2}
\sum_{p,q}
x^p_{\scriptscriptstyle S}
\left[G_{\scriptscriptstyle S}^{-1}\right]_{pq} x^q_{\scriptscriptstyle S}
-
\frac{1}{2}
\sum_{r,p}
\left(
x^p_{\scriptscriptstyle S}
-
\sum_{q}
\Bigl[
{G_{\scriptscriptstyle D}}G_{\scriptscriptstyle S}^{-1}
\Bigr]_{pq}
x^q_{\scriptscriptstyle S}
\right)
\Bigl[
\left(
{G_{\scriptscriptstyle S}}-{G_{\scriptscriptstyle D}}
\right)^{-1}\Bigr]_{rp}x^r_{\scriptscriptstyle S}
$$
\begin{equation} 
\mbox{}
-
\frac{1}{2}
\sum_{p,q}
x^p_{\scriptscriptstyle S}
\left[G_{\scriptscriptstyle S}^{-1}\right]_{pq} \gamma_q
-
\frac{1}{2}
\sum_{p}
\left(
x^p_{\scriptscriptstyle S}
+
\sum_{q}
\Bigl[
{G_{\scriptscriptstyle D}}G_{\scriptscriptstyle S}^{-1}
\Bigr]_{pq}
x^q_{\scriptscriptstyle S}
\right)
\sum_r
\Bigl[
\left(
{G_{\scriptscriptstyle S}}-{G_{\scriptscriptstyle D}}
\right)^{-1}
\Bigr]_{rp}
\gamma_r
.\end{equation} 
Next, we will need to use the relations:
\begin{equation} 
\sum_q\left[G_{\scriptscriptstyle S}^{-1}\right]_{pq} \gamma_q
=
\sum_q
\left[\left({G_{\scriptscriptstyle S}}-{G_{\scriptscriptstyle D}}\right)^{-1}\right]_{pq} 
\left(
x^q_{\scriptscriptstyle D}
-
\sum_{r}
\Bigl[
{G_{\scriptscriptstyle D}}G_{\scriptscriptstyle S}^{-1}
\Bigr]_{qr}
x^r_{\scriptscriptstyle S}
\right)
,\end{equation} 
$$
\sum_{q}
\Bigl[
\left(
{G_{\scriptscriptstyle S}}-{G_{\scriptscriptstyle D}}
\right)^{-1}
\Bigr]_{pq}\gamma_q
=
$$
\begin{equation} 
\mbox{}
=
\sum_q
\Bigl[
\left(
I-{G_{\scriptscriptstyle D}}G_{\scriptscriptstyle S}^{-1}
\right)^{-1}
\Bigr]_{pq}
\biggl(
\sum_{q}
\Bigl[\left({G_{\scriptscriptstyle S}}-{G_{\scriptscriptstyle D}}\right)^{-1}\Bigr]_{pq}
x^q_{\scriptscriptstyle D}
-
\sum_{q}
\Bigl[\left({G_{\scriptscriptstyle S}}-{G_{\scriptscriptstyle D}}\right)^{-1}\Bigr]_{pq}
\sum_{r}
\Bigl[
{G_{\scriptscriptstyle D}}G_{\scriptscriptstyle S}^{-1}
\Bigr]_{qr}
x^r_{\scriptscriptstyle S}
\biggr)
.\end{equation} 
As a result, we obtain the following expression for the parameter of the light beam $\gamma_1$:
$$
\gamma_1
=
x^1_{\scriptscriptstyle S}
+
\frac{1}{2}
\sum_{p,q}
x^p_{\scriptscriptstyle S}
\left[G_{\scriptscriptstyle S}^{-1}\right]_{pq} x^q_{\scriptscriptstyle S}
-
\frac{1}{2}
\sum_{p,q}
\left(
x^p_{\scriptscriptstyle S}
-
\sum_{r}
\Bigl[
{G_{\scriptscriptstyle D}}G_{\scriptscriptstyle S}^{-1}
\Bigr]_{pr}
x^r_{\scriptscriptstyle S}
\right)
\sum_r
\Bigl[
\left(
{G_{\scriptscriptstyle S}}-{G_{\scriptscriptstyle D}}
\right)^{-1}\Bigr]_{rp}x^r_{\scriptscriptstyle S}
$$
$$
\mbox{}
-
\frac{1}{2}
\sum_{p,q}
x^p_{\scriptscriptstyle S}
\left[\left({G_{\scriptscriptstyle S}}-{G_{\scriptscriptstyle D}}\right)^{-1}\right]_{pq} 
\left(
x^q_{\scriptscriptstyle D}
-
\sum_{r}
\Bigl[
{G_{\scriptscriptstyle D}}G_{\scriptscriptstyle S}^{-1}
\Bigr]_{qr}
x^r_{\scriptscriptstyle S}
\right)
$$
$$
\mbox{}
-
\frac{1}{2}
\sum_{p,q}
\left(
x^p_{\scriptscriptstyle S}
+
\sum_{r}
\Bigl[
{G_{\scriptscriptstyle D}}G_{\scriptscriptstyle S}^{-1}
\Bigr]_{pr}
x^r_{\scriptscriptstyle S}
\right)
\Bigl[
\left(
I-{G_{\scriptscriptstyle D}}G_{\scriptscriptstyle S}^{-1}
\right)^{-1}
\Bigr]_{pq}
\biggl(
\sum_{r}
\Bigl[\left({G_{\scriptscriptstyle S}}-{G_{\scriptscriptstyle D}}\right)^{-1}\Bigr]_{pr}
x^r_{\scriptscriptstyle D}
$$
\begin{equation} 
\mbox{}
-
\sum_{s,r}
\Bigl[\left({G_{\scriptscriptstyle S}}-{G_{\scriptscriptstyle D}}\right)^{-1}\Bigr]_{ps}
\Bigl[
{G_{\scriptscriptstyle D}}G_{\scriptscriptstyle S}^{-1}
\Bigr]_{sr}
x^r_{\scriptscriptstyle S}
\biggr)
\label{gamma1end}
.\end{equation} 
Similarly, from the relations (\ref{kappaPfirst}) taking into account the relations (\ref{gammaPend}) for $\gamma_p$ we can find the required form of parameters $k_p/k_1$:
$$
\frac{k_p}{k_1}=
\sum_{q}\left[G_{\scriptscriptstyle S}^{-1}\right]_{pq}x^q_{\scriptscriptstyle S}
-
\sum_{q,s}
\left[G_{\scriptscriptstyle S}^{-1}\right]_{pq}
\Bigl[
\left(
I-{G_{\scriptscriptstyle D}}G_{\scriptscriptstyle S}^{-1}
\right)^{-1}
\Bigr]_{qs}
\left(
x^s_{\scriptscriptstyle D}
-
\sum_{r}
\Bigl[
{G_{\scriptscriptstyle D}}G_{\scriptscriptstyle S}^{-1}
\Bigr]_{sr}
x^r_{\scriptscriptstyle S}
\right)
=
$$
$$
\mbox{}
=
\sum_{q}\left[G_{\scriptscriptstyle S}^{-1}\right]_{pq}x^q_{\scriptscriptstyle S}
+
\sum_{q}
\Bigl[
\left(
{G_{\scriptscriptstyle D}}
-
G_{\scriptscriptstyle S}^{-1}
\right)^{-1}
\Bigr]_{pq}
\left(
x^q_{\scriptscriptstyle D}
-
\sum_{r}
\Bigl[
{G_{\scriptscriptstyle D}}G_{\scriptscriptstyle S}^{-1}
\Bigr]_{qr}
x^r_{\scriptscriptstyle S}
\right)
=
$$
$$
\mbox{}
=
\sum_{q}
\Bigl[
\left(
{G_{\scriptscriptstyle D}}
-
G_{\scriptscriptstyle S}^{-1}
\right)^{-1}
\Bigr]_{pq}
x^q_{\scriptscriptstyle D}
+
\sum_{q,r}
x^r_{\scriptscriptstyle S}
\left(
\delta_r^q
\left[G_{\scriptscriptstyle S}^{-1}\right]_{pq}
-
\Bigl[
\left(
{G_{\scriptscriptstyle D}}
-
G_{\scriptscriptstyle S}^{-1}
\right)^{-1}
\Bigr]_{pq}
\Bigl[
{G_{\scriptscriptstyle D}}G_{\scriptscriptstyle S}^{-1}
\Bigr]_{qr}
\right)
=
$$
\begin{equation} 
\mbox{}
=
\sum_{q}
\Bigl[\left({G_{\scriptscriptstyle D}}-G_{\scriptscriptstyle S}^{-1}\right)^{-1}\Bigr]_{pq}
x^q_{\scriptscriptstyle D}
+
\sum_{q}
\left(
\left[G_{\scriptscriptstyle S}^{-1}\right]_{pq}
-
\Bigl[\left({G_{\scriptscriptstyle S}}-{G_{\scriptscriptstyle S}}G_{\scriptscriptstyle D}^{-1}G_{\scriptscriptstyle S}^{-1}\right)^{-1}\Bigr]_{pq}
\right)
x^q_{\scriptscriptstyle S}
\label{kappaPend}
.\end{equation} 

Note that the expressions obtained above for the parameters of the light signal can be formally redefined using the obtained retarded time equation for the light beam (\ref{RetardedTimeEq}), which gives an additional equation linking the coordinates of the source and the detector.
In this case, the numerical values of the parameters will not change.

%

\begin{adjustwidth}{-\extralength}{0cm}

\reftitle{References}

\PublishersNote{}
\end{adjustwidth}

\begin{thebibliography}{999}

\bibitem[Abbott et~al.(2016)Abbott, Abbott, Abbott, Abernathy, Acernese, and
  et~al.]{PhysRevLett.116.061102}
Abbott, B.P.; Abbott, R.; Abbott, T.D.; Abernathy, M.R.; Acernese, F.; et~al..
\newblock Observation of Gravitational Waves from a Binary Black Hole Merger.
\newblock {\em Phys. Rev. Lett.} {\bf 2016}, {\em 116},~061102.
\newblock {\url{https://doi.org/10.1103/PhysRevLett.116.061102}}.

\bibitem[Abbott et~al.(2019)Abbott, Abbott, Abbott, Abraham, Acernese, and
  et~al.]{PhysRevX.9.031040}
Abbott, B.P.; Abbott, R.; Abbott, T.D.; Abraham, S.; Acernese, F.; et~al..
\newblock {GWTC-1}: A Gravitational-Wave Transient Catalog of Compact Binary
  Mergers Observed by {LIGO} and {Virgo} during the First and Second Observing
  Runs.
\newblock {\em Phys. Rev. X} {\bf 2019}, {\em 9},~031040.
\newblock {\url{https://doi.org/10.1103/PhysRevX.9.031040}}.

\bibitem[Abbott et~al.(2021)Abbott, Abbott, Abraham, Acernese, Ackley, and
  et~al.]{PhysRevX.11.021053}
Abbott, R.; Abbott, T.D.; Abraham, S.; Acernese, F.; Ackley, K.; et~al..
\newblock {GWTC-2}: Compact Binary Coalescences Observed by {LIGO} and {Virgo}
  during the First Half of the Third Observing Run.
\newblock {\em Phys. Rev. X} {\bf 2021}, {\em 11},~021053.
\newblock {\url{https://doi.org/10.1103/PhysRevX.11.021053}}.

\bibitem[Odintsov and Oikonomou(2024)]{ODINTSOV2024101562}
Odintsov, S.; Oikonomou, V.
\newblock The necessity of multi-band observations of the stochastic
  gravitational wave background.
\newblock {\em Physics of the Dark Universe} {\bf 2024}, {\em 46},~101562.
\newblock {\url{https://doi.org/https://doi.org/10.1016/j.dark.2024.101562}}.

\bibitem[{EPTA Collaboration and InPTA Collaboration:} et~al.(2023){EPTA
  Collaboration and InPTA Collaboration:}, {Antoniadis, J.}, {Arumugam, P.},
  {Arumugam, S.}, {Babak, S.}, and {Bagchi, M. et al.}]{AandArefId0}
{EPTA Collaboration and InPTA Collaboration:}.; {Antoniadis, J.}.; {Arumugam,
  P.}.; {Arumugam, S.}.; {Babak, S.}.; {Bagchi, M. et al.}.
\newblock The second data release from the {European} Pulsar Timing Array -
  {III}. Search for gravitational wave signals.
\newblock {\em Astronomy \& Astrophysics} {\bf 2023}, {\em 678},~A50.
\newblock {\url{https://doi.org/10.1051/0004-6361/202346844}}.

\bibitem[Reardon et~al.(2023)Reardon, Zic, Shannon, Hobbs, and {Matthew Bailes
  et al.}]{Reardon_2023}
Reardon, D.J.; Zic, A.; Shannon, R.M.; Hobbs, G.B.; {Matthew Bailes et al.}.
\newblock Search for an Isotropic Gravitational-wave Background with the
  {Parkes} Pulsar Timing Array.
\newblock {\em The Astrophysical Journal Letters} {\bf 2023}, {\em 951},~L6.
\newblock {\url{https://doi.org/10.3847/2041-8213/acdd02}}.

\bibitem[Xu et~al.(2023)Xu, Chen, Guo, Jiang, and {Bojun Wang et al.}]{Xu_2023}
Xu, H.; Chen, S.; Guo, Y.; Jiang, J.; {Bojun Wang et al.}.
\newblock Searching for the Nano-Hertz Stochastic Gravitational Wave Background
  with the {Chinese} Pulsar Timing Array Data Release {I}.
\newblock {\em Research in Astronomy and Astrophysics} {\bf 2023}, {\em
  23},~075024.
\newblock {\url{https://doi.org/10.1088/1674-4527/acdfa5}}.

\bibitem[Thorne(1980)]{Thorne:1980ru}
Thorne, K.S.
\newblock {Multipole Expansions of Gravitational Radiation}.
\newblock {\em Rev. Mod. Phys.} {\bf 1980}, {\em 52},~299--339.
\newblock {\url{https://doi.org/10.1103/RevModPhys.52.299}}.

\bibitem[Baumgarte and Shapiro(1998)]{Baumgarte:1998te}
Baumgarte, T.W.; Shapiro, S.L.
\newblock {On the numerical integration of Einstein's field equations}.
\newblock {\em Phys. Rev. D} {\bf 1998}, {\em 59},~024007,
  \href{http://arxiv.org/abs/gr-qc/9810065}{{\normalfont [gr-qc/9810065]}}.
\newblock {\url{https://doi.org/10.1103/PhysRevD.59.024007}}.

\bibitem[Blanchet(2014)]{Blanchet:2013haa}
Blanchet, L.
\newblock {Gravitational Radiation from Post-Newtonian Sources and Inspiralling
  Compact Binaries}.
\newblock {\em Living Rev. Rel.} {\bf 2014}, {\em 17},~2,
  \href{http://arxiv.org/abs/1310.1528}{{\normalfont [arXiv:gr-qc/1310.1528]}}.
\newblock {\url{https://doi.org/10.12942/lrr-2014-2}}.

\bibitem[Mukhanov et~al.(1992)Mukhanov, Feldman, and
  Brandenberger]{MUKHANOV1992203}
Mukhanov, V.; Feldman, H.; Brandenberger, R.
\newblock Theory of cosmological perturbations.
\newblock {\em Physics Reports} {\bf 1992}, {\em 215},~203--333.
\newblock {\url{https://doi.org/https://doi.org/10.1016/0370-1573(92)90044-Z}}.

\bibitem[Ma and Bertschinger(1995)]{Ma19957}
Ma, C.P.; Bertschinger, E.
\newblock Cosmological perturbation theory in the synchronous and conformal
  {Newtonian} gauges.
\newblock {\em Astrophysical Journal} {\bf 1995}, {\em 455},~7--25.
\newblock {\url{https://doi.org/10.1086/176550}}.

\bibitem[Landau and Lifshitz(1975)]{LandauEng1}
Landau, L.D.; Lifshitz, E.M.
\newblock {\em The Classical Theory of Fields}, 4th ed.; Vol.~2, {\em Course of
  Theoretical Physics Series}, Butterworth-Heinemann: Oxford(UK),  1975; p.
  402.

\bibitem[Bennett et~al.(2013)Bennett, Larson, Weiland, Jarosik, and Hinshaw~et
  al.]{Bennett2013}
Bennett, C.; Larson, D.; Weiland, J.; Jarosik, N.; Hinshaw~et al., G.
\newblock Nine-year {Wilkinson} Microwave Anisotropy Probe {(WMAP)}
  observations: Final maps and results.
\newblock {\em Astrophysical Journal, Supplement Series} {\bf 2013}, {\em 208}.
\newblock {\url{https://doi.org/10.1088/0067-0049/208/2/20}}.

\bibitem[Secrest et~al.(2021)Secrest, von Hausegger, Rameez, Mohayaee, Sarkar,
  and Colin]{Secrest_2021}
Secrest, N.J.; von Hausegger, S.; Rameez, M.; Mohayaee, R.; Sarkar, S.; Colin,
  J.
\newblock A Test of the Cosmological Principle with Quasars.
\newblock {\em The Astrophysical Journal Letters} {\bf 2021}, {\em 908},~L51.
\newblock {\url{https://doi.org/10.3847/2041-8213/abdd40}}.

\bibitem[{Siewert, Thilo M.} et~al.(2021){Siewert, Thilo M.}, {Schmidt-Rubart,
  Matthias}, and {Schwarz, Dominik J.}]{refId0}
{Siewert, Thilo M.}.; {Schmidt-Rubart, Matthias}.; {Schwarz, Dominik J.}.
\newblock Cosmic radio dipole: Estimators and frequency dependence.
\newblock {\em Astronomy and Astrophysics} {\bf 2021}, {\em 653},~A9.
\newblock {\url{https://doi.org/10.1051/0004-6361/202039840}}.

\bibitem[Mittal et~al.(2023)Mittal, Oayda, and Lewis]{10.1093/mnras/stad3706}
Mittal, V.; Oayda, O.T.; Lewis, G.F.
\newblock {The cosmic dipole in the Quaia sample of quasars: a Bayesian
  analysis}.
\newblock {\em Monthly Notices of the Royal Astronomical Society} {\bf 2023},
  {\em 527},~8497--8510,
  \href{http://arxiv.org/abs/https://academic.oup.com/mnras/article-pdf/527/3/8497/54645877/stad3706.pdf}{{\normalfont
  [https://academic.oup.com/mnras/article-pdf/527/3/8497/54645877/stad3706.pdf]}}.
\newblock {\url{https://doi.org/10.1093/mnras/stad3706}}.

\bibitem[Osetrin et~al.(2006)Osetrin, Obukhov, and Filippov]{OsetrinHomog2006}
Osetrin, K.E.; Obukhov, V.V.; Filippov, A.E.
\newblock Homogeneous spacetimes and separation of variables in the
  {H}amilton--{J}acobi equation.
\newblock {\em Journal of Physics A: Mathematical and General} {\bf 2006}, {\em
  39},~6641--6647.
\newblock {\url{https://doi.org/10.1088/0305-4470/39/21/S64}}.

\bibitem[Bagrov and Obukhov(1983)]{Obukhov1983}
Bagrov, V.G.; Obukhov, V.V.
\newblock Classes of Exact Solutions of the {E}instein--{M}axwell Equations.
\newblock {\em Annalen der Physik} {\bf 1983}, {\em 495},~181--188.
\newblock {\url{https://doi.org/10.1002/andp.19834950402}}.

\bibitem[Bagrov et~al.(1986)Bagrov, Obukhov, and Shapovalov]{Obukhov1986}
Bagrov, V.G.; Obukhov, V.V.; Shapovalov, A.V.
\newblock Special {S}täckel electrovac spacetimes.
\newblock {\em Pramana} {\bf 1986}, {\em 26},~93--108.
\newblock {\url{https://doi.org/10.1007/BF02847629}}.

\bibitem[Obukhov(2021)]{Obukhov2021695}
Obukhov, V.
\newblock Solutions of {Maxwell}’s Equations in Vacuum for {Stäckel} Spaces
  of Type (1.1).
\newblock {\em Russian Physics Journal} {\bf 2021}, {\em 64},~695--703.
\newblock {\url{https://doi.org/10.1007/s11182-021-02372-9}}.

\bibitem[Obukhov(2022)]{ObukhovSym14122595}
Obukhov, V.V.
\newblock {Maxwell} Equations in Homogeneous Spaces with Solvable Groups of
  Motions.
\newblock {\em Symmetry} {\bf 2022}, {\em 14}.
\newblock {\url{https://doi.org/10.3390/sym14122595}}.

\bibitem[Obukhov(2020)]{Obukhov20202050186}
Obukhov, V.V.
\newblock Separation of variables in {H}amilton--{J}acobi equation for a
  charged test particle in the {S}t{\"{a}}ckel spaces of type (2.1).
\newblock {\em International Journal of Geometric Methods in Modern Physics}
  {\bf 2020}, {\em 17}.
\newblock {\url{https://doi.org/10.1142/S0219887820501868}}.

\bibitem[Obukhov(2022)]{ObukhovUniverse8040245}
Obukhov, V.V.
\newblock {Maxwell's} Equations in Homogeneous Spaces for Admissible
  Electromagnetic Fields.
\newblock {\em Universe} {\bf 2022}, {\em 8}.
\newblock {\url{https://doi.org/10.3390/universe8040245}}.

\bibitem[Obukhov and Kartashov(2024)]{Obukhov2024_Einstein-Maxwell}
Obukhov, V.; Kartashov, D.
\newblock Einstein-{Maxwell} Equations for Homogeneous Spaces.
\newblock {\em Russ. Phys. J} {\bf 2024}, {\em 67},~193--197.
\newblock {\url{https://doi.org/10.1007/s11182-024-03108-1}}.

\bibitem[Obukhov(2023)]{ObukhovSym15030648}
Obukhov, V.V.
\newblock Exact Solutions of {Maxwell} Equations in Homogeneous Spaces with the
  Group of Motions {G3(VIII)}.
\newblock {\em Symmetry} {\bf 2023}, {\em 15}.
\newblock {\url{https://doi.org/10.3390/sym15030648}}.

\bibitem[Osetrin et~al.(2022{\natexlab{a}})Osetrin, Osetrin, and
  Osetrina]{Osetrin2022EPJP856}
Osetrin, K.; Osetrin, E.; Osetrina, E.
\newblock Geodesic deviation and tidal acceleration in the gravitational wave
  of the {Bianchi} type {IV} universe.
\newblock {\em European Physical Journal Plus} {\bf 2022}, {\em 137}.
\newblock {\url{https://doi.org/10.1140/epjp/s13360-022-03061-3}}.

\bibitem[Osetrin et~al.(2022{\natexlab{b}})Osetrin, Osetrin, and
  Osetrina]{Osetrin2022894}
Osetrin, K.; Osetrin, E.; Osetrina, E.
\newblock Gravitational wave of the {Bianchi} {VII} universe: particle
  trajectories, geodesic deviation and tidal accelerations.
\newblock {\em European Physical Journal C} {\bf 2022}, {\em 82}.
\newblock {\url{https://doi.org/10.1140/epjc/s10052-022-10852-6}}.

\bibitem[Osetrin et~al.(2023)Osetrin, Osetrin, and
  Osetrina]{Osetrin325205JPA_2023}
Osetrin, K.E.; Osetrin, E.K.; Osetrina, E.I.
\newblock Deviation of geodesics and particle trajectories in a gravitational
  wave of the {Bianchi} type {VI} universe.
\newblock {\em Journal of Physics A: Mathematical and Theoretical} {\bf 2023},
  {\em 56},~325205.
\newblock {\url{https://doi.org/10.1088/1751-8121/ace6e3}}.

\bibitem[Nojiri and Odintsov(2007)]{Odintsov2007}
Nojiri, S.; Odintsov, S.D.
\newblock Introduction to modified Gravity and gravitational alternative for
  dark energy.
\newblock {\em International Journal of Geometric Methods in Modern Physics}
  {\bf 2007}, {\em 04},~115--145.
\newblock {\url{https://doi.org/10.1142/S0219887807001928}}.

\bibitem[Nojiri and Odintsov(2011)]{Odintsov2011}
Nojiri, S.; Odintsov, S.D.
\newblock Unified cosmic history in modified gravity: {F}rom {F}({R}) theory to
  {L}orentz non-invariant models.
\newblock {\em Physics Reports} {\bf 2011}, {\em 505},~59--144.
\newblock {\url{https://doi.org/10.1016/j.physrep.2011.04.001}}.

\bibitem[Capozziello and {De Laurentis}(2011)]{Capozziello2011}
Capozziello, S.; {De Laurentis}, M.
\newblock Extended Theories of Gravity.
\newblock {\em Physics Reports} {\bf 2011}, {\em 509},~167--321.
\newblock {\url{https://doi.org/10.1016/j.physrep.2011.09.003}}.

\bibitem[Nojiri et~al.(2017)Nojiri, Odintsov, and Oikonomou]{Odintsov2017}
Nojiri, S.; Odintsov, S.D.; Oikonomou, V.K.
\newblock Modified gravity theories on a nutshell: {I}nflation, bounce and
  late-time evolution.
\newblock {\em Physics Reports} {\bf 2017}, {\em 692},~1--104.
\newblock
  {\url{https://doi.org/https://doi.org/10.1016/j.physrep.2017.06.001}}.

\bibitem[Odintsov and Oikonomou(2022)]{Odintsov:2022hxu}
Odintsov, S.D.; Oikonomou, V.K.
\newblock {Chirality of gravitational waves in Chern-Simons f(R) gravity
  cosmology}.
\newblock {\em Phys. Rev. D} {\bf 2022}, {\em 105},~104054,
  \href{http://arxiv.org/abs/2205.07304}{{\normalfont
  [arXiv:gr-qc/2205.07304]}}.
\newblock {\url{https://doi.org/10.1103/PhysRevD.105.104054}}.

\bibitem[Elizalde et~al.(2024)Elizalde, Nojiri, Odintsov, and
  Oikonomou]{Elizalde:2023rds}
Elizalde, E.; Nojiri, S.; Odintsov, S.D.; Oikonomou, V.K.
\newblock {Propagation of gravitational waves in a dynamical wormhole
  background for two-scalar Einstein\textendash{}Gauss\textendash{}Bonnet
  theory}.
\newblock {\em Phys. Dark Univ.} {\bf 2024}, {\em 45},~101536,
  \href{http://arxiv.org/abs/2312.02889}{{\normalfont
  [arXiv:gr-qc/2312.02889]}}.
\newblock {\url{https://doi.org/10.1016/j.dark.2024.101536}}.

\bibitem[Nojiri et~al.(2024)Nojiri, Odintsov, and Oikonomou]{Nojiri:2023mbo}
Nojiri, S.; Odintsov, S.D.; Oikonomou, V.K.
\newblock {Propagation of gravitational waves in Einstein-Gauss-Bonnet gravity
  for cosmological and spherically symmetric spacetimes}.
\newblock {\em Phys. Rev. D} {\bf 2024}, {\em 109},~044046,
  \href{http://arxiv.org/abs/2311.06932}{{\normalfont
  [arXiv:gr-qc/2311.06932]}}.
\newblock {\url{https://doi.org/10.1103/PhysRevD.109.044046}}.

\bibitem[Odintsov et~al.(2023)Odintsov, Oikonomou, Giannakoudi, Fronimos, and
  Lymperiadou]{sym15091701}
Odintsov, S.D.; Oikonomou, V.K.; Giannakoudi, I.; Fronimos, F.P.; Lymperiadou,
  E.C.
\newblock Recent Advances in Inflation.
\newblock {\em Symmetry} {\bf 2023}, {\em 15}.
\newblock {\url{https://doi.org/10.3390/sym15091701}}.

\bibitem[Odintsov et~al.(2024)Odintsov, D'Onofrio, and Paul]{Odintsov:2024sbo}
Odintsov, S.D.; D'Onofrio, S.; Paul, T.
\newblock {Primordial gravitational waves in horizon cosmology and constraints
  on entropic parameters}.
\newblock {\em Phys. Rev. D} {\bf 2024}, {\em 110},~043539,
  \href{http://arxiv.org/abs/2407.05855}{{\normalfont
  [arXiv:gr-qc/2407.05855]}}.
\newblock {\url{https://doi.org/10.1103/PhysRevD.110.043539}}.

\bibitem[Osetrin et~al.(2021)Osetrin, Kirnos, Osetrin, and
  Filippov]{OsetrinSymmetry2021}
Osetrin, K.; Kirnos, I.; Osetrin, E.; Filippov, A.
\newblock Wave-Like Exact Models with Symmetry of Spatial Homogeneity in the
  Quadratic Theory of Gravity with a Scalar Field.
\newblock {\em Symmetry} {\bf 2021}, {\em 13}.
\newblock {\url{https://doi.org/10.3390/sym13071173}}.

\bibitem[Osetrin et~al.(2023)Osetrin, Kirnos, and
  Osetrin]{Osetrin356Universe_2023}
Osetrin, K.; Kirnos, I.; Osetrin, E.
\newblock An Exact Model of a Gravitational Wave in the {Bianchi} {III}
  Universe Based on {Shapovalov} {II} Wave Spacetime and the Quadratic Theory
  of Gravity.
\newblock {\em Universe} {\bf 2023}, {\em 9}.
\newblock {\url{https://doi.org/10.3390/universe9080356}}.

\bibitem[Osetrin et~al.(2024)Osetrin, Epp, and Chervon]{OSETRIN2024169619}
Osetrin, K.E.; Epp, V.Y.; Chervon, S.V.
\newblock Propagation of light and retarded time of radiation in a strong
  gravitational wave.
\newblock {\em Annals of Physics} {\bf 2024}, {\em 462},~169619.
\newblock {\url{https://doi.org/https://doi.org/10.1016/j.aop.2024.169619}}.

\bibitem[Shapovalov(1978{\natexlab{a}})]{Shapovalov1978I}
Shapovalov, V.N.
\newblock Symmetry and separation of variables in {Hamilton}-{Jacobi} equation.
  {I}.
\newblock {\em Soviet Physics Journal} {\bf 1978}, {\em 21},~1124--1129.
\newblock {\url{https://doi.org/10.1007/BF00894559}}.

\bibitem[Shapovalov(1978{\natexlab{b}})]{Shapovalov1978II}
Shapovalov, V.N.
\newblock Symmetry and separation of variables in {Hamilton}-{Jacobi} equation.
  {II}.
\newblock {\em Soviet Physics Journal} {\bf 1978}, {\em 21},~1130--1132.
\newblock {\url{https://doi.org/10.1007/BF00894560}}.

\bibitem[Shapovalov(1979)]{Shapovalov1979}
Shapovalov, V.N.
\newblock The {St{\"{a}}ckel} spaces.
\newblock {\em Sib. Math. Journal (Sov. J. of Math.)} {\bf 1979}, {\em
  20},~790--800.
\newblock {\url{https://doi.org/10.1007/BF00971844}}.

\bibitem[Obukhov and Osetrin(2004)]{Obukhov:2004yek}
Obukhov, V.V.; Osetrin, K.E.
\newblock {Variables separation in gravity}.
\newblock {\em PoS} {\bf 2004}, {\em WC2004},~027,
  \href{http://arxiv.org/abs/gr-qc/0502033}{{\normalfont [gr-qc/0502033]}}.
\newblock {\url{https://doi.org/10.22323/1.013.0027}}.

\bibitem[Osetrin and Osetrin(2020)]{Osetrin2020Symmetry}
Osetrin, K.; Osetrin, E.
\newblock Shapovalov wave-like spacetimes.
\newblock {\em Symmetry} {\bf 2020}, {\em 12}.
\newblock {\url{https://doi.org/10.3390/SYM12081372}}.

\end{thebibliography}
\end{document}